\documentclass[seceq,preprint]{ptptex}

\usepackage{graphicx}
\input{amssym.def}
\input{amssym.tex}
\renewcommand{\thefootnote}{\fnsymbol{footnote}}
\preprintnumber[3cm]{%<-- [..]: optional width of preprint # column.
hep-th/0312122 \\
UT-03-39\\ December, 2003}
\title{%        %You can use \\ for explicit line-break
A Universal Nonlinear Relation among\\
Boundary States in Closed String Field Theory
}

\author{%       %Use \scshape  for the family name
Isao \textsc{Kishimoto},\footnote{
E-mail:~ikishimo@hep-th.phys.s.u-tokyo.ac.jp}
Yutaka \textsc{Matsuo}\footnote{
E-mail:~matsuo@phys.s.u-tokyo.ac.jp}
and
Eitoku \textsc{Watanabe}\footnote{
E-mail:~eytoku@hep-th.phys.s.u-tokyo.ac.jp}
}

\inst{%         %Affiliation, neglected when [addenda] or [errata]
Department of Physics, University of Tokyo, Tokyo 113-0033, Japan
}

\abst{%         %this abstract is neglected when [addenda] or [errata]
We show that the boundary states satisfy a
nonlinear relation (the idempotency equation) with respect
to the star product of closed string field theory.
This relation is universal in the sense that various D-branes,
including the infinitesimally deformed ones,
satisfy the same equation, {\em including} the coefficient.
This paper generalizes our analysis (hep-th/0306189)
in the following senses. (1) We present
a background-independent
formulation based on conformal field theory.
It illuminates the geometric nature of the relation
and 
allows us to more systematically
 analyze the variations around the D-brane
background.
(2) We show that the Witten-type star product satisfies
a similar relation but with a more divergent coefficient.
(3) We determine the coefficient of the relation
analytically. The result shows that the $\alpha$ parameter can
be formally factored out, and the relation becomes universal.
We present a conjecture on vacuum theory based on this
computation.
}

\begin{document}

\maketitle

\setcounter{footnote}{0}
\renewcommand{\thefootnote}{\arabic{footnote}}

\section{Introduction}
In quantum field theories, soliton solutions play
an essential role in providing an understanding of the non-perturbative
properties of the system. 
Well-known examples are the instanton and monopole
solutions in non-abelian gauge theories and the black hole
solutions in Einstein gravity.
They are exact solutions of nonlinear equations
that reflect the topological structure of the configuration
space. In string theory, the corresponding objects are
D-branes and the NS 5-brane. The discoveries of these solutions
have led to major breakthroughs in the history of string theory.

Among these solitons, D-branes \cite{r:Polchin}
are known to have  completely
stringy descriptions in terms of boundary states \cite{r:Boundary}.
They include all the effects of the massive modes of the
closed string in a compact form.
A natural question is whether D-branes can be understood
as solutions of a nonlinear equation
of string field theory (SFT), as in particle theory.

Such a question was originally
posed  in the context of open string
field theory \cite{r:Witten}.
In particular, in the proposal of vacuum string
field theory (VSFT) \cite{r:VSFT} (see also the partial list
of Refs.~\citen{r:others,r:RZa,r:projectors,r:butterfly} for
related studies),
it is conjectured that D-branes are described by
projectors of Witten's star
product.  This idea has stimulated widespread
interest in string field theory.
It seems to be a natural idea in the context
of noncommutative geometry, where D-branes are described
by noncommutative solitons, namely the projection operators
\cite{r:GMS}.

There exists, however, a technical challenge in this
scenario.  The boundary state belongs
to the closed string Hilbert space.  In order to describe
it with the {\em open} string variable, 
it is necessary to use singular states, such as sliver
or butterfly states.
The treatment of these states is usually subtle,
and one needs a careful analysis including
a regularization scheme.\footnote{See Ref.~\citen{r:butterfly} for
the recent studies based on butterfly states
with a regularized treatment of the mid-point singularity.}

In a previous paper \cite{r:kmw1}, we proposed to look at this problem
from a different angle.
We used closed string field theory directly
and attempted to derive nonlinear relations that
are satisfied by  the  boundary states.
In particular, we examined the star product for a closed string
(the covariant version of the light-cone vertex)
studied by Hata, Itoh, Kugo, Kunitomo and Ogawa (HIKKO)
\cite{r:HIKKO2}
and proved that boundary states
satisfy the idempotency equation
\begin{equation}\label{eq_idempotency}
|B\rangle \star |B\rangle \propto
|B\rangle\,,
\end{equation}
which is strikingly  similar to the conjecture \cite{r:VSFT}
in the open string case.
We have confirmed the validity of 
this identity for a large family of boundary states,
including those with constant electro-magnetic flux.
Furthermore, variations of this equation seem to give proper
on-shell conditions of the open string mode on the D-brane.

The computation in Ref.~\citen{r:kmw1} is based on the oscillator
formulation \cite{r:HIKKO2}.
That computation has the definite merit that it is reduced to
straightforward algebraic manipulations. At the same time, however,
it has certain drawbacks, such as divergence resulting from
the infinite dimensionality of the Neumann matrices.
It also depends explicitly on the flat
background where the closed string oscillator is defined.

In this paper, we use an alternative formulation of
string field theory based on conformal field theory.
This is  the formulation
which was developed by LeClair, Peskin and Preitskopf
(LPP)  \cite{r:LPP}.
It sheds light on some issues that are obscure in
the oscillator formulation.  For example, because only
the properties of the stress-energy tensor are used,
it does not depend on the particular representation
of the background CFT.
It also provides a clear geometrical picture,
and sometimes we can define a natural prescription
of the regularization
in which the oscillator calculation becomes ambiguous
due to divergences.

In particular, we are able to prove the idempotency relation
in a background-independent fashion.
We also derive some explicit results 
answering questions that were left
unanswered in the previous paper.  Specifically,
we compute (a) the normalization factor in the relation
(\ref{eq_idempotency}).  We show in particular that the
unphysical $\alpha$ parameters can be factored out of
this relation.  As an important corollary, we show that
the idempotency relation is {\em universal}, including
the coefficient for any boundary states in the flat background.
This result is essential to develop a possible vacuum theory.
(b) For the deformation of the boundary state through a 
vector-type variation,
which corresponds to a massless photon on the D-brane,
we show that transversality is implied from (\ref{eq_idempotency}).
We could not reach this conclusion in Ref.~\citen{r:kmw1}, as it
contains a term of the form $0\times \infty$ in the coefficient,
making it difficult to treat.
(c) We examine the star product
using a Witten-type three string vertex
for the closed string \cite{r:Z_rev}.  We show, in particular,
that the basic relation (\ref{eq_idempotency}) remains valid
while the coefficient in the relation
diverges very strongly, as $\delta(0)^\infty$.
This divergence occurs whenever we consider an overlap of
the boundary states.
We note that a similar divergence appears in the inner product
of boundary states.
In any case, our analysis shows that the idempotency relation is valid
for both HIKKO and Witten type vertices.  This is another sort
of the universality of the idempotency relation.

We organize the paper as follows.  In \S 2, we give a review of
the HIKKO theory in LPP language. In \S 3, we re-express the main
statements of Ref.~\citen{r:kmw1} in this formulation.  We also include
several new results, which are proved in the later sections.
Readers who are unfamiliar with Ref.~\citen{r:kmw1} may wish to
read this section
first to understand the overall picture of our results.
In \S 4, we  derive  the relation
(\ref{eq_idempotency}) for a Witten-type vertex,
which is used in nonpolynomial closed string field theory
\cite{r:Z_rev,r:SZ,r:KKS,r:KS}.
In \S 5, we give a background-independent proof
of the idempotency relation in the conformal field theory picture.
In \S 6, we show that the solutions to the infinitesimal deformation
of the relation
around a boundary state
can be identified with the open string spectrum on the D-brane.
In the appendices, we explain our notation
and the correspondence with our previous paper.
 We also give an explicit computation of the coefficient
appearing in the star product.
%

%%%%%%%%%%%%%%%%%%%%%%%%%%%%%%%%%%%%%%%%%%%%%%%%%%%%%%%%%%%%%%
\section{CFT description of closed string field theory}
\label{sec_LPP}
%%%%%%%%%%%%%%%%%%%%%%%%%%%%%%%%%%%%%%%%%%%%%%%%%%%%%%%%%%%%%%

\subsection{String vertices from conformal mapping}
In string field theory, the $N$-string vertex $\langle v_N|$, which
specifies the interactions of $N$ strings,
is the  fundamental object to construct the action.
$\langle v_N|$ is a mapping from $N$ string Hilbert space to the set of
complex numbers: $\mathcal{H}^{\otimes N} \rightarrow {\mathbf{C}}$.
LeClair, Peskin and Preitskopf  (LPP) \cite{r:LPP} defined  $\langle v_N|$
in terms of $N$ conformal mappings $h_r(w_r)$ (with $r=1,\cdots,N$)
from $N$ disks with coordinates $w_r$ to a Riemann sphere $\Sigma$.
For each element in the Hilbert space
$|A_r\rangle\in \mathcal{H},~r=1,\cdots, N$,
we denote by ${\cal O}_{A_r}(w_r)$ the corresponding operator
defined through $|A_r\rangle={\cal O}_{A_r}(0)|0\rangle$.
LeClair, Peskin and Preitskopf defined  $\langle v_N|$ using these
data as
\begin{eqnarray}
\label{eq:LPPvertex}
 \langle v_N|A_1\rangle|A_2\rangle\cdots |A_N\rangle
=\langle h_1[{\cal O}_{A_1}]\,h_2[{\cal O}_{A_2}]
\cdots h_N[{\cal O}_{A_N}]\rangle\,,
\end{eqnarray}
where the right-hand side is a correlation function of
conformal field theory (CFT) on $\Sigma$. 
Here, $h_r[{\cal O}_{A_r}]$ is an operator on $\Sigma$
defined by applying the conformal transformation $h_r$
to the operator ${\cal O}_{A_r}$.
If $\mathcal{O}(w_r)$  is a primary field
of conformal dimension $h$, the image of the mapping is 
$h_r[\mathcal{O}(0)]=(dh_r(w_r)/dw_r)^h\mathcal{O}(h_r(w_r))|_{w_r=0}$.
The anti-holomorphic part is given in the same way.

In the HIKKO formulation of closed string field theory,
there is an extra parameter (the $\alpha$-parameter),
 which represents the length of the closed string
at the interaction point.
This is an analogous to the light-cone momentum $p^+$ in the
light-cone string field theory \cite{r:Mandel, r:C-G, r:LCG}.
It is additively conserved during the entire process.
We need to include the dependence
on $\alpha$ in the string field, and we make this explicit by writing
\begin{eqnarray}
 |\Phi(\alpha)\rangle=|\Phi\rangle\otimes|\alpha\rangle\,,
\end{eqnarray}
where the ket vector $|\alpha\rangle$ is the eigenvector of the operator
$\hat{\alpha}$ with eigenvalue $\alpha$.
The normalization of eigenstates is given by
$\langle \alpha_1|\alpha_2\rangle=2\pi\delta(\alpha_1-\alpha_2)$.
The other factor, $|\Phi\rangle$, is an element of
the conventional string field, which can be expanded as
\begin{eqnarray}
\label{eq:expand_Phi}
 |\Phi\rangle=\sum_{A}{\cal O}_A|0\rangle \phi_A\,\,,
\end{eqnarray}
where $\phi_A$ is a component field and ${\cal O}_A|0\rangle$ is
a closed string state.

We define two kinds of products between string fields, the
dot ($\cdot$) product and the star (${\star}$) product, using CFT
language.
The dot product of two string fields yields a complex number (i.e., 
$\mathcal{H}\otimes \mathcal{H}\rightarrow \mathbf{C}$):
\begin{eqnarray}
\label{eq:dot_Phi12}
 \Phi_1(\alpha_1)\cdot \Phi_2(\alpha_2)
&\equiv &-2\pi\delta(\alpha_1+\alpha_2)(-1)^{|\Phi_1|}
\langle \Phi_1| b_0^-|\Phi_2\rangle\nonumber\\
&\equiv &-2\pi\delta(\alpha_1+\alpha_2)(-1)^{|\Phi_1|}
\langle I[\Phi_1]\,b_0^-\Phi_2\rangle\\
&=&(-1)^{|\Phi_1||\Phi_2|}\Phi_2(\alpha_2)\cdot \Phi_1(\alpha_1)\,,
\label{eq_commutativity}
\end{eqnarray}
where
$I(z)=1/z$ is the inversion map,
$\langle \cdots \rangle$ represents a correlator of CFT on $\Sigma$
and $(-1)^{|\Phi|}$ denotes the Grassmann parity of the string field
$\Phi$. In addition to the $\alpha$-dependent factor,
we insert $b_0^-$ into the correlator as a convention.
The last equality implies that the dot product satisfies
the (graded) commutativity Eq.~(\ref{eq_commutativity}).
We can rewrite Eq.~(\ref{eq:dot_Phi12}) using the reflector
$\langle \hat{R}(1,2)|$ as
\begin{eqnarray}
\Phi_1(\alpha_1)\cdot \Phi_2(\alpha_2)
= \langle \hat{R}(1,2)|b_0^{-(2)}|\Phi_1\rangle_1|\Phi_2\rangle_2\,,
\end{eqnarray}
where $\langle \hat{R}(1,2)|$
is obtained from the LPP reflector\footnote{
In the following, we use the term ``the LPP vertex''
in reference to the vertex operator
defined by Eq.~(\ref{eq:LPPvertex}) without inclusion of
the ghost zero-mode insertion and the state vectors
for the $\alpha$ parameter.}
 (2-string vertex) $\langle R(1,2)|$ given in Eq.~(\ref{eq:refl_LPP}):
\begin{eqnarray}
 \langle \hat{R}(1,2)|=-\int {d\alpha_1\over 2\pi}{d\alpha_2\over 2\pi}\,
{}_1\langle \alpha_1|{}_2\langle \alpha_2|\langle R(1,2)|
2\pi\delta(\alpha_1+\alpha_2)\,.
\end{eqnarray}

The ${\star}$ product, which determines the 3-string vertex,
is defined by the CFT correlator,\footnote{
In Ref.~\citen{r:HIKKO2}, the dot product
and the star product are defined using purely oscillator
language, and their explicit correspondence to the LPP language can be
understood from Ref.~\citen{r:Kunitomo-Suehiro}.
We summarize this correspondence in Appendix \ref{sec:osc-LPP}.
Note that our convention is
different from that used in Ref.~\citen{r:Z_rev}.
} in combination with the dot product:
\begin{eqnarray}
\label{eq:Phi^3}
(\Phi_1(\alpha_1)\star\Phi_2(\alpha_2))\cdot\Phi_3(\alpha_3)
&=&-2\pi\delta(\alpha_1+\alpha_2+\alpha_3)(-1)^{|\Phi_2|}\nonumber\\
&&\times \left\langle h_1[b_0^{-}\wp\Phi_1]\,
h_2[b_0^{-}\wp\Phi_2]\,
h_3[b_0^{-}\wp\Phi_3]\,
\right\rangle\\
&=&(-1)^{|\Phi_1|(|\Phi_2|+|\Phi_3|)}
(\Phi_2(\alpha_2)\star\Phi_3(\alpha_3))\cdot\Phi_1(\alpha_1)
\label{eq:cyclic}
\\
&=&\Phi_1(\alpha_1)\cdot (\Phi_2(\alpha_2)\star\Phi_3(\alpha_3))\,.
\end{eqnarray}
We note that an operator $b_0^-\wp$ is inserted into the correlator.
The projector $\wp$ is defined by
\begin{eqnarray}
\label{eq:levelmatch}
 \wp=\int^{\pi}_{-\pi}{d\theta\over 2\pi}e^{i\theta(L_0-\tilde{L}_0)}\,,
\end{eqnarray}
where $L_0$ and $\tilde{L}_0$ are total (i.e., matter$+$ghost) Virasoro
operator.
It imposes the level matching condition on string fields.

\begin{figure}[htbp]
	\begin{center}
	\scalebox{0.6}[0.6]{\includegraphics{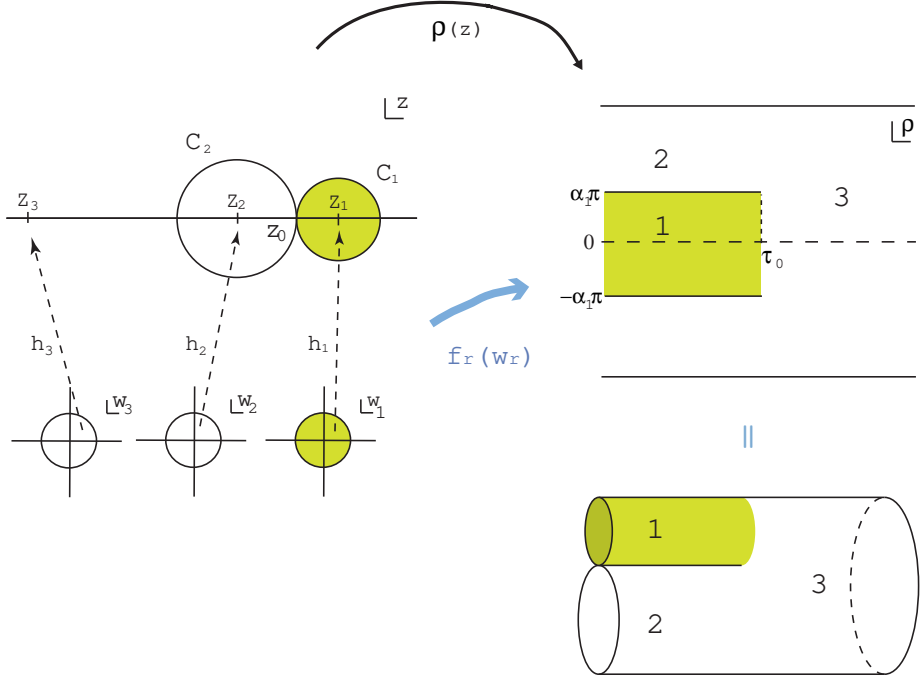}}
	\end{center}
	\caption{$\rho(z)$ represents Mandelstam map.
The images of string 1 and 2 at the interaction time $\tau_0$ are
the contours $C_1$ and $C_2$, respectively. $C_1+C_2$ becomes string 3.}
	\label{fig:Mandel}
\end{figure}

The conformal mappings $h_r(w_r)$ (with $r=1,2,3$), which represent
the overlapping configuration,
are given in terms of the Mandelstam map \cite{r:Mandel}
from the complex plane with coordinate $z$ into the
(pants-shaped) $\rho$-plane
(Fig.~\ref{fig:Mandel}):
\begin{equation}
 \rho(z)=\alpha_1\log(z-1)+\alpha_2\log z\,.
\end{equation}
The mapping $h_r(w_r)$ is defined by combining a logarithmic function
that maps the $r$-th disk to the $r$-th strip in the $\rho$-plane and
the inverse of Mandelstam map:
\begin{eqnarray}
\label{eq:hrdef}
h_r(w_r)=\rho^{-1}(f_r(w_r))\,,\quad
f_r(w_r)=\alpha_r\log w_r+\tau_0
+i\beta_r\,.
\end{eqnarray}
In the logarithmic function $f_r(w_r)$,
the argument of $w_r$ should be taken from $-\pi$ to $\pi$.
We divide it into two regions (corresponding to
positive and negative argument) to define the parameter
$\beta_r={\rm sgn}(\mbox{arg}(w_r))\pi\sum_{s=1}^{r-1}\alpha_s$.
The interaction time $\tau_0$ is defined as
$ \tau_0={\rm Re}\,\rho(z_0)=\sum_{r=1}^3\alpha_r\log|\alpha_r|$
($\alpha_3=-\alpha_1-\alpha_2$),
where
$z_0=-\alpha_2/\alpha_3$ is the solution to $d\rho(z)/dz=0$.
The cyclic symmetry (\ref{eq:cyclic}) can be demonstrated
by the $SL(2,R)$ transformation in the $z$-plane.
The 3-string vertex $\langle \hat{V}(1,2,3)|$ is given by
\begin{eqnarray}
(\Phi_1(\alpha_1) \star \Phi_2(\alpha_2))\cdot \Phi_3(\alpha_3)=
\langle \hat{V}(1,2,3)|b_0^{-(3)}
|\Phi_1(\alpha_1)\rangle_1|\Phi_2(\alpha_2)\rangle_2
|\Phi_3(\alpha_3)\rangle_3\,.~~~
\end{eqnarray}
The inserted operator $b_0^{-(3)}$ here comes from the dot product.
{}From Eq.~(\ref{eq:Phi^3}), this vertex can be expressed in terms of
the LPP vertex (\ref{eq:LPPvertex}) and some factors as
\begin{eqnarray}
\label{eq:HIKKOv3LPP}
\langle \hat{V}(1,2,3)|b_0^{-(3)}&=&-\int {d\alpha_1\over 2\pi}
{d\alpha_2\over 2\pi}{d\alpha_3\over 2\pi}\,
{}_1\langle\alpha_1|{}_2\langle\alpha_2|{}_3\langle\alpha_3|\,
{}_{123}\langle v_3|\\
&&\times\,
b_0^{-(1)}\wp^{(1)}\,b_0^{-(2)}\wp^{(2)}\,b_0^{-(3)}\wp^{(3)}.
\nonumber
\end{eqnarray}

The basic properties of the $\star$ product are summarized as
\begin{eqnarray}
\label{eq:anti-com}
&&\Phi_1\star\Phi_2=-(-1)^{|\Phi_1||\Phi_2|}\Phi_2\star\Phi_1
\,\,,
\\
\label{eq:Jacobi}
&& \Phi_1 \star (\Phi_2 \star \Phi_3)
+(-1)^{|\Phi_1|(|\Phi_2|+|\Phi_3|)}\Phi_2 \star (\Phi_3 \star\Phi_1)
\nonumber\\
&&~~~~
+(-1)^{|\Phi_3|(|\Phi_1|+|\Phi_2|)}\Phi_3 \star (\Phi_1 \star\Phi_2)=0\,\,.
\end{eqnarray}
The first line [``(anti-)commutativity'']
follows from Eq.~(\ref{eq:Phi^3}) and the $SL(2,C)$-invariance
of the conformal vacuum.
The second line [the ``Jacobi identity''] is much more nontrivial
and we need to use the critical dimension $d=26$  to prove it
\cite{r:HIKKO2,r:AKT2}.\footnote{
Originally, this identity was shown using the
Cremmer-Gervais identity \cite{r:C-G} in the oscillator formulation.
To prove this identity in terms of LPP language, the generalized gluing
and re-smoothing theorem (GGRT) is essential.
\cite{r:LPP2,r:SenS,r:AKT1}
}
We note that this identity should not be confused with associativity,
which generally does not hold for the closed string star product.

\subsection{Action}
In this subsection, we give a brief review of the HIKKO
closed string field theory.
We note, however, that the definition of the star product alone
is sufficient to understand our claims.  For this reason,
the reader may skip to the next section to see our basic claims.
We use the HIKKO construction to explain the ghost zero-mode
convention of the physical state and
to investigate an analogy with VSFT.

The action of the HIKKO closed string field theory is similar
to that of Witten's open string field theory,
\begin{eqnarray}
\label{eq:action}
 S={1\over 2}\Phi\cdot Q_B\Phi+{1\over 3}g\,\Phi\cdot (\Phi \star \Phi)\,,
\end{eqnarray}
where $Q_B=\oint j_B+\oint \tilde{j}_B$ is the conventional
nilpotent BRST operator for a closed string:
\begin{eqnarray}
\label{eq:BRST}
&& Q_B=c_0^+(L_0+\tilde{L}_0-2)+{1\over 2}c_0^-(L_0-\tilde{L}_0)\\
&&~~~~~~~~+(M+\tilde{M})b_0^+ +2(M-\tilde{M})b_0^- + Q_B'\,,\nonumber\\
&&M=-\sum_{n=1}^{\infty}nc_{-n}c_n\,,~~~~~
\tilde{M}=-\sum_{n=1}^{\infty}n\tilde{c}_{-n}\tilde{c}_n\,,\\
&&Q_B'=\sum_{n\ne 0}\left(c_{-n}L_n^{\rm matter}
+\tilde{c}_{-n}\tilde{L}_n^{\rm matter}
\right)\nonumber\\
&&~~~~~~~~+\sum_{m,n,m+n\ne 0}{m-n\over 2}\left(c_m c_nb_{-m-n}+
\tilde{c}_m \tilde{c}_n\tilde{b}_{-m-n}\right).
\end{eqnarray}
We have written the dependence on ghost zero modes explicitly 
for later convenience.
The operator $Q_B$ has the following properties with respect to
the dot product and the ${\star}$ product.
For the  dot product, we have
\begin{eqnarray}
\label{eq:partial}
 (Q_B\Phi_1(\alpha_1))\cdot \Phi_2(\alpha_2)
&=&-(-1)^{|\Phi_1|}\Phi_1(\alpha_1)\cdot(Q_B\Phi_2(\alpha_2))\\
&&
-\pi\delta(\alpha_1+\alpha_2)\langle I[\Phi_1](L_0-\tilde{L}_0)\Phi_2\rangle
\nonumber
\end{eqnarray}
from Eq.~(\ref{eq:dot_Phi12}), using contour deformation
in the CFT correlator and the relation
$\{b_0^-,Q_B\}={1\over 2}(L_0-\tilde{L}_0)$.
In particular, if $\Phi_1$ or $\Phi_2$ satisfies the level matching
condition, the second term on the right-hand side  vanishes,
and therefore the ``partial integration formula'' holds.
We can show that the BRST charge $Q_B$ is a
derivation with respect to the ${\star}$ product:
\begin{eqnarray}
\label{eq:Q_Bderivation}
 Q_B(\Phi_1\star\Phi_2)=(Q_B\Phi_1) \star \Phi_2
+(-1)^{|\Phi_1|}\Phi_1 \star (Q_B\Phi_2)\,.
\end{eqnarray}
Here we have used Eq.~(\ref{eq:partial}),
$(L_0-\tilde{L}_0)|\Phi_1\star\Phi_2\rangle=0$,
the contour deformation in Eq.~(\ref{eq:Phi^3}) and anti-commutativity
(i.e., the relation $\{Q_B,b_0^-\wp\}=0$).

In the action Eq.~(\ref{eq:action}), the string fields
are subject to some constraints:
(1) the  string field $|\Phi\rangle$
should have ghost number 3 (i.e.,
each ${\cal O}_A$ in the expansion Eq.~(\ref{eq:expand_Phi})
is a ghost number 3 operator);
(2) $|\Phi\rangle$ has odd Grassmann parity;
(3)  we impose the reality condition in the sense that we have
\begin{eqnarray}
\label{eq:reality}
(|\Phi\rangle_2)^{\dagger}=\langle \hat{R}(1,2)|\Phi\rangle_1\,;
\end{eqnarray}
(4) the level matching condition
$(L_0-\tilde{L}_0)|\Phi\rangle=0$
(or  $\wp|\Phi\rangle=|\Phi\rangle$) is imposed. 

The action Eq.~(\ref{eq:action}) is invariant
under nonlinear gauge transformations of the string field.
Thus we have
\begin{eqnarray}
 \delta_{\Lambda}\Phi=Q_B\Lambda+g(\Phi \star\Lambda-\Lambda \star\Phi)\,,
\end{eqnarray}
where $|\Lambda\rangle$ is a gauge parameter that
has ghost number 2 and even Grassmann parity
and satisfies the level matching condition.
We can easily confirm the gauge invariance
$\delta_{\Lambda}S=0$ using the nilpotency $Q_B^2=0$ and
the properties of dot and star products 
(\ref{eq:dot_Phi12}),~(\ref{eq:Phi^3}),~(\ref{eq:anti-com}),~(\ref{eq:Jacobi}).

By expanding the string field $\Phi$ with respect to the ghost zero mode,
\begin{eqnarray}
\label{eq:field_expansion}
 |\Phi\rangle=c_0^-|\phi\rangle+c_0^-c_0^+|\psi\rangle
+|\chi\rangle+c_0^+|\eta\rangle\,,
\end{eqnarray}
the kinetic term of the action (\ref{eq:action})
becomes
\begin{eqnarray}
\label{eq:kinetic}
\langle I[\Phi]b_0^-Q_B\Phi\rangle
=\langle I[\phi](L_0+\tilde{L}_0-2)\phi\rangle +\cdots\,,
\end{eqnarray}
where we have omitted the $\alpha$-dependent factor.
This shows that the physical sector is contained in
the slot  $c_0^-|\phi\rangle$ in the expansion
 Eq.~(\ref{eq:field_expansion}) in our convention.
In fact,  the gauge in which $\psi=\chi=\eta=0$
is adopted in Ref.~\citen{r:HIKKO2}
in order to obtain the gauge fixed action from the gauge invariant
action (\ref{eq:action}).
By contrast, we remove the (internal) ghost number constraint from
 $c_0^-|\phi\rangle$ to include FP ghosts in SFT.

\section{A universal nonlinear relation for the boundary state}

\subsection{Summary of our previous results}

In our previous paper  \cite{r:kmw1}, we use
the boundary state of the D$p$-brane
in a flat background with a constant electro-magnetic
flux $F_{\mu\nu}$:\footnote{We note that the notation for the
ghost fields used here is slightly different from that in
our previous paper \cite{r:kmw1}. The correspondence between
them is explained in Appendix \ref{sec:osc-LPP}.}
\begin{eqnarray}\label{eq_boundary_state}
   |B(x^\perp)\rangle &=& \exp\left({-\sum_{n=1}^\infty{1\over n}
\alpha_{-n}\mathcal{O}\tilde{\alpha}_{-n}
+\sum_{n=1}^\infty (c_{-n}\tilde b_{-n}
+\tilde c_{-n} b_{-n})}
\right)\\
&&\times\, c_0^+c_1\tilde{c}_1|p^{\parallel}=0,x^{\perp}\rangle
\otimes |0\rangle_{gh},\nonumber\\
\mathcal{O}^{\mu}_{~\nu} & = & \left[(1+F)^{-1}(1-F)\right]^{\mu}_{~\nu}\,,
\quad \mu,\nu=0,1,\cdots,p\,,\\
\mathcal{O}^i_{~j} & = & -\delta^i_j\,,\quad
~~~~~~~~~~~~~~~~~~~~~~~i,j=p+1,\cdots,d-1\,.
\end{eqnarray}
Here, $p^\parallel$ (resp., $x^\perp$) is the momentum (resp., coordinate)
along the Neumann (resp., Dirichlet) directions.
The ghost sector is fixed by the boundary conditions
$(c_n+\tilde c_{-n})|B\rangle_{gh}=
(b_n-\tilde b_{-n})|B\rangle_{gh}=0$.
The state $|0\rangle_{gh}$ is the $SL(2,C)$ invariant vacuum.
For the matter sector, the relation
$(L_n^{\rm{matter}}-\tilde{L}^{\rm
matter}_{-n})|B(x^{\perp})\rangle=0$ is satisfied,
because ${\cal O}$ in the exponent is an orthogonal matrix.
This implies $Q_B|B(x^{\perp})\rangle=0$ for the conventional
BRST operator $Q_B$ (\ref{eq:BRST}).

We need to slightly modify
the boundary state to follow the convention of the string fields
in the previous section \cite{r:kmw1}. Here, we define
\begin{equation}
\label{eq:Phi_B}
   |\Phi_B(x^\perp,\alpha)\rangle \equiv
c_0^- b_0^+ |B(x^\perp)\rangle \otimes |\alpha\rangle\,.
\end{equation}
We include the $\alpha$ parameter here
to define the $\star$ product and the factor
 $c_0^- b_0^+$.
The ghost factor $c_0^- b_0^+$ has the effect
of replacing $c_0^+$ with $c_0^-$,
so that the state is placed
in the correct slot of the string field (\ref{eq:field_expansion}).
The first nontrivial statement in Ref.~\citen{r:kmw1} is that
this modified boundary state satisfies the idempotency relation
for $\alpha_1\alpha_2>0$,
\begin{equation}\label{eq_idempotency2}
   \Phi_B(x^\perp,\alpha_1)\star \Phi_B(y^\perp,\alpha_2)
   = \delta^{d-p-1}(x^\perp-y^\perp) \,\mathcal{C}\, c_0^+
\Phi_B(y^\perp,\alpha_1+\alpha_2)\,.
\end{equation}
The constant factor $\mathcal{C}$ is given formally in terms
of the determinant of an infinite-dimensional matrix
and was not fixed analytically in Ref.~\citen{r:kmw1}.
It turns out that it can be determined
by using the so-called Cremmer-Gervais identity, \cite{r:C-G}
which is explained in Appendix \ref{sec_determinant}.
It takes a very simple form only for the critical
dimension, $d=26$, in which case we have
\begin{equation}
   \mathcal{C}= K^3 |\alpha_1\alpha_2(\alpha_1+\alpha_2)|\,,
\end{equation}
where $K$ is an infinite constant that depends on the cutoff.
It is universal in the sense that it is independent of
$F_{\mu\nu}$, $x^{\perp}$ and $p$.
Therefore, if we change the normalization of $\Phi_B$ so that
$\tilde\Phi(x^\perp,\alpha)\equiv \frac{-1}{g_o|\alpha|} 
\Phi_B(x^\perp,\alpha)$,
together with the inclusion of the usual open
string coupling ($g_o=g^{1/2}$),
the dependence on the parameter $\alpha$
formally drops from the idempotency relation,
\begin{equation}\label{eq_idempotency4}
   \delta^{d-p-1}(x^\perp-y^\perp) \mathcal{Q} \tilde\Phi(x^\perp,
   \alpha_1+\alpha_2) +g'
\tilde\Phi(x^\perp,\alpha_1)\star  \tilde\Phi(y^\perp,\alpha_2)=0\,\,.
\end{equation}
Here, $\mathcal{Q}\equiv \hat\alpha^2 c_0^+$ is
the ``pure ghost BRST operator,''
which we discuss below, and
$g'=g_o K^{-3}$ is the renormalized string coupling constant.

If we wish to interpret (\ref{eq_idempotency4}) as
an equation of motion of a possible vacuum theory,
the appearance of the delta function is annoying, as
it depends on the number of the transverse directions
explicitly.  It can be removed, however, if
we take a superposition of D$p$-brane
boundary states
along the transverse direction,
\begin{equation}
\Phi_f(\alpha)\equiv \int d^{d-p-1}
x^\perp f(x^\perp) \tilde\Phi(x^\perp,
\alpha)\,.
\end{equation}
Suppose $f$ satisfies the equation
\begin{equation}
\label{eq:projector}
f^2(x^\perp)=f(x^\perp)\,,
\end{equation}
Then, the relation for $\Phi_f$ can be written in
the universal form
\begin{equation}
\label{eq:eom}
  \mathcal{Q}\Phi_f(\alpha_1+\alpha_2)+g'\Phi_f(\alpha_1)
\star \Phi_f(\alpha_2)=0\,.
\end{equation}
We remark that the constraint (\ref{eq:projector})
has the form of a ``noncommutative'' soliton
for the commutative ring of functions in the transverse direction.\footnote{
The algebraic constraint
for $f$  is satisfied by  $f(x^\perp)=1$  (if $x^\perp \in \Sigma$)
and $f(x^\perp)=0$ (otherwise) for some subset
$\Sigma$ of $\mathbf{R}^{d-p-1}$.  
This fixing of $f$ describes the distribution
of D-branes in the transverse direction.
The discrete nature of $f$ is
regularized in the noncommutative space-time, namely by including
a $B$ field along the transverse direction \cite{r:SW}.
In this case, the equation for
$f$ should be changed to  $f* f=f$, where $*$ is 
a product that reflects the noncommutativity.
This is similar to the equation of the noncommutative soliton
\cite{r:GMS}.
We hope to come back to this issue in a future paper.
 \label{fn:Moyal}}

The second nontrivial observation made in Ref.~\citen{r:kmw1} is
the deformation of (\ref{eq_idempotency2}).
We considered deformations of the type
\begin{equation}
\label{eq_deform}
   \delta\Phi_B(x^\perp,
\alpha) =\oint \frac{d\sigma}{2\pi} V(\sigma) \Phi_B(x^\perp,\alpha)\,.
\end{equation}
This describes infinitesimal deformations of the
boundary condition. It can be physically interpreted
as describing the infinitesimally curved D-brane
induced by the collective modes of the open strings.
In order that the variation preserves the idempotency relation
(\ref{eq_idempotency2}), we need to impose the condition
\begin{eqnarray}
\label{eq_idempotency3}
&& \delta \Phi_B(x^\perp, \alpha_1)\star \Phi_B(y^\perp,\alpha_2)
+  \Phi_B(x^\perp, \alpha_1)\star \delta \Phi_B(y^\perp,\alpha_2)
\nonumber\\
&&~~~~~~=  \delta^{d-p-1}(x^\perp-y^\perp)
   \mathcal{C}c_0^+ \delta \Phi_B(x^\perp, \alpha_1+\alpha_2)\,.
\end{eqnarray}
We examined the two simplest cases, in which
$V(\sigma)$ is given by (1) the scalar-type deformation
$V_S(\sigma)=:e^{ik_\mu X^\mu(\sigma)}:$
and (2) the vector-type deformation 
$V_V(\sigma)=:\zeta_\nu \partial_\sigma X^\nu e^{ik_\mu X^\mu(\sigma)}:$.
We have proved that the relation  (\ref{eq_idempotency3})
requires the on-shell conditions for open string tachyon and
vector particle,
\begin{eqnarray}
&& k_\mu G^{\mu\nu} k_\nu={1\over \alpha'}\qquad {\mbox{for }} V=V_S\,,\qquad
  k_\mu G^{\mu\nu} k_\nu=0\qquad {\mbox{for }} V=V_V\,,
\end{eqnarray}
where
$$ G^{\mu\nu} \equiv \left[(1+F)^{-1}\eta (1-F)\right]^{\mu\nu}$$
is the ``open string metric'' \cite{r:SW}.

We also demonstrated that the vector-type variation has a
``gauge symmetry'' of the form $\zeta_\nu=\zeta_\nu+\epsilon k_\nu$,
owing to the relation
\begin{equation}
   \oint \frac{d\sigma}{2\pi}(k_\nu \partial_\sigma X^\nu)
e^{ik_\mu X^\mu(\sigma)}\Phi_B=-i \oint
\frac{d\sigma}{2\pi}\partial_\sigma\left(e^{ik_\mu X^\mu(\sigma)}\right)
\Phi_B=0\,.
\end{equation}
On the other hand, the requirement of the transversality
condition
\begin{equation}
   \zeta\cdot k\equiv \zeta_\mu G^{\mu\nu} k_\nu=0
\end{equation}
is rather subtle, because we encounter a coefficient of the
form $0\times \infty$ multiplying this factor.
While the appearance of such a subtlety is inevitable
in the operator formalism, we prove in \S \ref{sec:spectrum} 
that a regularized expression
is obtained in the CFT approach and that the transversality
condition is indeed needed.

To summarize, in all the examples we studied,
the deformation of the idempotency relation precisely
reproduces the spectrum of an open string.
This convincingly shows that Eq.~(\ref{eq_idempotency2})
is a very good characterization of D-branes,
including the infinitesimally curved branes.

\begin{figure}[htbp]
	\begin{center}
	\scalebox{0.7}[0.7]{\includegraphics{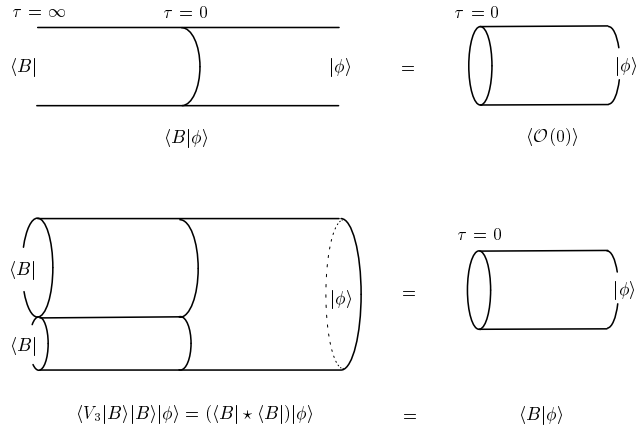}}
	\end{center}
	\caption{Star product of boundary states.}
	\label{fig:star_prod}
\end{figure}

We now give an intuitive proof of the idempotency relation
Eq.~(\ref{eq_idempotency2}).
We first explain the nature of the boundary
state as a surface state.  Consider an inner product between
a boundary state $|B\rangle$ and a vector in
the closed string Fock space $|\phi\rangle=\mathcal{O}|0\rangle$.
As is well-known, $\langle B|\phi\rangle$ gives
a one point function on a disk $\langle \mathcal{O}(0)\rangle$
with the boundary condition at the boundary $|w|=1$
specified by the boundary state.
By the conformal mapping $w=e^{\tau+i\sigma}$, the disk is mapped
to a half-infinite cylinder that is cut at $\tau=0$.
Therefore, as depicted in the first figure in Fig.~\ref{fig:star_prod},
the boundary state is characterized by two operations
as a surface state:
(1) to cut the infinite cylinder at $\tau=0$, and
(2) to set an appropriate boundary condition at the edge.

When we calculate the star product between the boundary
states
$(\langle B| \star \langle B|)|\phi\rangle$,
 we prepare a pants diagram which represents the HIKKO vertex
(the second figure of Fig.~\ref{fig:star_prod}),
and  we attach boundary states at its two legs.
As the surface states, they have the effect of stripping off the two
legs at the interaction time $\tau_0$
and set the same boundary conditions along the two circles.
Suppose we can ignore the curvature singularity at
the interaction point. Then we are left with a half infinite
cylinder with the boundary condition of $|B\rangle$.
This is in effect the same as taking the inner product
$\langle B|\phi\rangle$ (Fig.~\ref{fig:star_prod}).

{}From the above discussion, we see the uniqueness of the boundary state
with respect to the star product of the closed string.
{}From the nontrivial topology of the pants diagram,
it is difficult to imagine that anything other than
the boundary state can satisfy the idempotency relation.
In this sense, although it is difficult to prove rigorously,
the solution set of the idempotency relation
Eq.~(\ref{eq_idempotency2}) seems to be identical to
the set of consistent boundary states.

%%%%%%%%%%%%%%%%%%%%%%%%%%%%%%%
\subsection{Vacuum theory for closed string fields}
Since the idempotency relation (\ref{eq:eom}) for the boundary states
takes a  universal form,
it  may be natural to consider the action whose
equation of motion is given by this relation.
We now describe some properties of such a theory
in order to find an analogy to VSFT \cite{r:VSFT}.
This theory is very similar to the HIKKO theory, which
we reviewed in the previous section.
Consider an action of the form
\begin{eqnarray}
\label{eq:action_general}
  S={1\over 2}\Phi\cdot {\cal Q}\Phi
+{g'\over 3}\Phi\cdot (\Phi\star\Phi)\,,
\end{eqnarray}
where the operator ${\cal Q}=\hat{\alpha}^2c_0^+$
is that appearing in Eq.~(\ref{eq_idempotency4}).
It has ghost number 1 and satisfies the same type of relations as
the original BRST operator $Q_B$,
\begin{eqnarray}
\label{eq:generalQ}
&&{\cal Q}^2=0\,,~~~~~
({\cal Q}\Phi_1)\cdot \Phi_2
=-(-1)^{|\Phi_1|}\Phi_1\cdot {\cal Q}\Phi_2\,,\nonumber\\
&&{\cal Q}(\Phi_1 \star \Phi_2)=({\cal Q}\Phi_1)\star\Phi_2
+(-1)^{|\Phi_1|}\Phi_1\star({\cal Q}\Phi_2)\,.
\end{eqnarray}
The first equation here is obvious.
The last two identities can be proved using the relation
$c_0^+={1\over 2}\oint {dw\over 2\pi i}w^{-2}c(w)+
{1\over 2}\oint {d\bar{w}\over 2\pi i}\bar{w}^{-2}\tilde{c}(\bar{w})$
and Eqs.~(\ref{eq:dot_Phi12}) and (\ref{eq:Phi^3}).
Note that $\hat{\alpha}^2$ is necessary
to cancel the conformal factor.\footnote{
The same operator, $\hat{\alpha}^2c_0^+$,
  was also considered in Ref.~\citen{r:AKT2} in a different context.
We can also check these identities in the oscillator language
using properties of the Neumann coefficients.
}
They ensure the gauge invariance of the action under the transformation
\begin{eqnarray}
 \delta_{\Lambda}\Phi
={\cal Q}\Lambda+g'(\Phi \star\Lambda-\Lambda \star\Phi)\,.
\end{eqnarray}
The equation of motion for the action (\ref{eq:action_general}) is
\begin{eqnarray}
\label{eq:EOM}
 %{\cal Q}\Phi +g'\Phi\star\Phi=0\,.
b_0^-\left({\cal Q}\Phi +g'(\Phi\star\Phi)\right)=0\,.
\end{eqnarray}
This is very similar to the relation (\ref{eq:eom}).
Suppose that it can be extended to the
case $\alpha_1\alpha_2<0$.\footnote{
That such an extension would be possible seems
 natural from the cyclicity of the product (\ref{eq:cyclic}).
There exist, however, some subtleties similar to
the divergence of the norm of the boundary state.}
Formally, its solution is given by
\begin{equation}
\Phi_0(\tilde \alpha)=\lim_{M\rightarrow \infty}
\frac{1}{2M}\int_{-M}^{M} d\alpha\, e^{i\alpha\tilde \alpha}
\Phi_f(\alpha)\,.
\end{equation}
We note that the appearance of the Fourier transformation with respect
to the parameter $\alpha$ is typical in the HIKKO computation of string
amplitudes, \cite{r:HIKKO3} where we must impose the condition that all
the ``momentum'' in the external lines have the same $\tilde \alpha$.
The divergent normalization factor $(2M)^{-1}$ can be removed through
the renormalization of the coupling constant: $g' \rightarrow \tilde{g}=
g'(2M)^{-1}$. This situation is also parallel to that in
Ref.~\citen{r:HIKKO3}.

We now consider the re-expansion of 
the action (\ref{eq:action_general}) around the
nontrivial classical solution $\Phi_0(\tilde \alpha)$,
\begin{eqnarray}
\label{eq:S'}
 S'={1\over 2}\Phi\cdot {\cal Q}_0\Phi+{\tilde g\over 3}\Phi\cdot
(\Phi \star\Phi)+S_0\,,
\end{eqnarray}
where the new kinetic term ${\cal Q}_0$ is given by
\begin{eqnarray}
\label{eq:Q0}
 {\cal Q}_0\Phi={\cal Q}\Phi+\tilde g(\Phi_0(\tilde \alpha)
\star\Phi-(-1)^{|\Phi|}\Phi\star\Phi_0(\tilde \alpha))\,,
\end{eqnarray}
and $S_0=-{\tilde g\over 6}\Phi_0\cdot (\Phi_0\star\Phi_0)$
corresponds to  ``potential height.''
This new kinetic term of $S'$ given in Eq.~(\ref{eq:S'}) also satisfies
the three identities (\ref{eq:generalQ}),
and therefore it possesses gauge invariance.

The above consideration suggests that we can consider
the ``vacuum version'' of closed string field theory
as an analogue of the vacuum string field theory (VSFT)
of the Witten-type {\it open} string field theory \cite{r:VSFT}.
As in VSFT, this pure ghost BRST operator has no physical states.
As we have observed in Ref.~\citen{r:kmw1} and will
refine the results in the following, we might interpret that the
boundary state is a classical solution $\Phi_0$
to the equation of motion (\ref{eq:EOM}) with ${\cal
Q}=\hat{\alpha}^2c_0^+$
and ${\cal Q}_0$ given in Eq.~(\ref{eq:Q0}) has {\it open} string
spectrum on the D-brane. This is strikingly similar to VSFT scenario
\cite{r:VSFT}.

We note that the string coupling $\tilde{g}$ is equal to
the open string coupling $g_{\rm open}=g_{\rm closed}^{1/2}$
up to an infinite constant factor.
This may be related to the fact that there seems to be no
physical closed string sector in $\mathcal{Q}_0$,
as far as we studied.  In this sense, the vacuum theory
that we are considering may be regarded as a purely
open string field theory, while we are using the closed
string variables and thus the properties of the star product
are very different.

Finally, we comment that there may be possibilities other
than considering the vacuum theory.  We note that there is a
close analogy between Eq.~(\ref{eq:eom}) and the wedge state
algebra \cite{r:RZa} in open string field theory,
\begin{equation}
W_n \star W_m =W_{n+m-1}\,,\quad
W_n = (|0\rangle)^{n-1}_{\star},
\end{equation}
with the correspondence 
$\alpha \leftrightarrow (n-1)$.
This link becomes more precise if we take the large $n$ limit,
where we can regard the parameter $n$ (after rescaling) as a continuous
parameter. In this limit, the wedge state becomes the sliver state.   
We hope to come back to this analogy in the future
to elucidate the explicit link between our closed string formulation and
VSFT.  

%%%%%%%%%%%%%%%%%%%%%%%%%%%%%%%%%%%%%%
\section{Comments on the Witten-type vertex in nonpolynomial closed SFT}
%%%%%%%%%%%%%%%%%%%%%%%%%%%%%%%%%%%%%%
%
Here we briefly turn our attention to
another formulation that has been well-examined,
the non-polynomial closed string field theory, \cite{r:Z_rev,r:SZ,r:KKS,r:KS}
%\cite{r:SZ}\cite{r:KKS}\cite{r:KS}
and we discuss the idempotency equation of boundary states
in this context.

The non-polynomial closed SFT was constructed as
a direct extension of Witten's open string field theory to closed
strings.
It has the merit that it does not contain extra parameters.
On the other hand, it is known that
an infinite number of higher-order interaction terms
are necessary in the classical action in order for the theory to
cover the moduli space properly \cite{r:SZ,r:KKS}.
While a formal method to construct all vertices is  known
in terms of the LPP formulation \cite{r:KS},
it seems  that it is impractical to perform explicit computations
to all orders.

For this reason, we restrict ourselves to a 3-string vertex,
where a computation similar to that given in Ref.~\citen{r:kmw1} is
possible with the knowledge of Neumann coefficients of Witten's open
SFT.\cite{r:GJ12} The 3-string vertex $\langle V_W(1,2,3)|$ of
a nonpolynomial closed SFT is defined using the LPP vertex as
\begin{eqnarray}
\label{eq:nonpolyv3}
 \langle V_W(1,2,3)|=\langle v_{3W}|
b_0^{-(1)}\wp^{(1)}b_0^{-(2)}\wp^{(2)}b_0^{-(3)}\wp^{(3)}\,,
\end{eqnarray}
where $\langle v_{3W}|$ is the LPP 3-string vertex
(\ref{eq:LPPvertex})
with conformal mappings \sloppy
$ h_r(w_r)=h^{-1}\left(e^{{2\pi r \over 3}i}h(w_r)^{2\over 3}\right)
$ (where $r=1,2,3$ and $h(w):={1+iw\over 1-iw}$),
which realize the Witten-type overlapping.
These maps are identical to those appearing in the open string.
This implies that the Neumann coefficients of the (anti-) holomorphic part
of $\langle v_{3W}|$ coincide with those of Witten's open SFT.
We carried out the calculation of the star product
of the boundary states by replacing the
Neumann coefficients in a previous paper \cite{r:kmw1}
with those of the Witten theory.
The nonlinear relations for the Neumann coefficients essential in the
computation are actually the same:
\begin{eqnarray}
\label{eq:GJmatter}
&&\sum_{t,l}V^{rt}_{nl}V_{lm}^{ts}=\delta_{nm}\delta^{rs},~~
\sum_{t,l}V^{rt}_{nl}V_{l0}^{ts}=V_{n0}^{rs},~~
\sum_{t,l}V_{0l}^{rt}V_{l0}^{ts}=2V_{00}^{rs},\\
&&\sum_{t,l}X^{rt}_{nl}X^{ts}_{lm}=\delta_{nm}\delta^{rs},~~
\sum_{t,l}X^{rt}_{nl}X^{ts}_{l0}=X_{n0}^{rs}\,.
\end{eqnarray}
Therefore, the computation of the ``$\star$-product''
of the boundary states is parallel to that in
Ref.~\citen{r:kmw1}.\footnote{
Here, we have used the notation in Ref.~\citen{r:IK} for Neumann
coefficients.
For the matter sector, the relations (\ref{eq:GJmatter}) have the same 
form as those of the light-cone gauge SFT.\cite{r:Yoneya}
There is a difference, however.
In the present case, the rank of the matrices $(1-n^{(m)2}),(1-n^{(g)2})$
is half of their size, where we define
$n^{(m)}=\left(
\begin{array}[tb]{cc}
 -V^{11}& -V^{12}\\
-V^{21}&-V^{22}
\end{array}
\right)\!,
n^{(g)}=\left(
\begin{array}[tb]{cc}
-X^{11}&-X^{12}\\
-X^{21}&-X^{22}
\end{array}
\right)$. This causes extra divergent factor ${\bf c}$.
}
The result is
\begin{eqnarray}
\label{eq:idempotency_W}
\Phi_B(x^{\perp})\star \Phi_B(y^{\perp})
&=&{\det}^{-{d\over 2}}((1-(V^{33})^2))
\det(1-(X^{33})^2)\,{\bf c}\,\\
&&\times\,
\delta^{d-p-1}(x^{\perp}-y^{\perp})\,
c_0^+b_0^-\Phi_B(x^{\perp})\,,\nonumber
\end{eqnarray}
where we have defined
$|\Phi_B(x^{\perp})\rangle=c_0^-b_0^+|B(x^{\perp})\rangle$,
which is the same form as Eq.~(\ref{eq:Phi_B}), while it does not contain
the $\alpha$-sector.\footnote{
We define the ``$\star$-product'' by
$\langle V_W(1,2,3)|\Phi_1\rangle_1|\Phi_2\rangle_2
={}_3\langle \Phi_1\star\Phi_2|$.
The extra $b_0^-$ factor on the right-hand side comes from
the difference between the conventions of the
HIKKO (\ref{eq:HIKKOv3LPP}) and the non-polynomial formulation
(\ref{eq:nonpolyv3}) in the three string vertex.
}
Up to a constant prefactor,
this nonlinear equation has the same form as the HIKKO-type
vertex (\ref{eq_idempotency2}).

\begin{figure}[htbp]
	\begin{center}
	\scalebox{0.3}[0.3]{\includegraphics{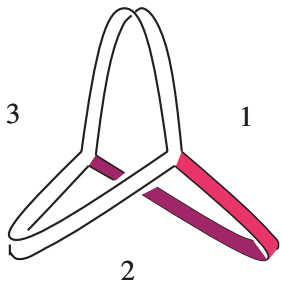}}
	\end{center}
	\caption{The halves of strings 1 and 2
overlap each other in the ``$\star$-product.''}
	\label{fig:Witten}
\end{figure}

In this case, however, the prefactor ${\bf c}$ in the above equation is
strongly divergent.  Specifically, there appears a factor of the form
$\prod_{\sigma\in \mbox{\scriptsize overlap}} \delta(0)$, which
should be regularized somehow.
This divergence results from the geometrical nature
of the Witten-type vertex (Fig.~{\ref{fig:Witten}}).
The problem is that it attaches two boundary states
point-wise for half of the boundary.
Roughly speaking, the boundary state contains a factor of
the form
$\prod_\sigma \delta(X(\sigma)-\tilde X(\sigma))$,
because it identifies the left and right movers.
On the other hand, the vertex operator contains the factor
$\prod_{\sigma, \sigma'} \delta(X^{(1)}(\sigma) -X^{(2)}(\sigma'))
 \delta(\tilde X^{(1)}(\sigma) -\tilde X^{(2)}(\sigma'))$,
 where $\sigma$ and $\sigma'$ are the coordinates
of the attached points on each string.
 If we take the star product of two boundary states,
we are left with $\delta(0)$ for each point $\sigma$ where
two boundary states are attached.

In the HIKKO $\star$ product (in the case $\alpha_1\alpha_2>0$),
no such divergent factor exists, because the overlapping part of strings
1 and 2 is only a point, i.e. the interaction point
($z_0$ in Fig.~{\ref{fig:Mandel}}).
This makes the HIKKO vertex appropriate for our purpose.

%%%%%%%%%%%%%%%%%%%%%%%%%%%%%%%%%%%%%%%%%%%%%%%%%%%%%%%%%%%%%%
\section{Proof of idempotency as a background-independent relation}
\label{sec_idempotency}
%%%%%%%%%%%%%%%%%%%%%%%%%%%%%%%%%%%%%%%%%%%%%%%%%%%%%%%%%%%%%%
%

In this section, we give a proof
of the idempotency relation of boundary states
using the CFT technique described in the preceding sections.
Compared with the proof carried out
in Ref.~\citen{r:kmw1} in terms of oscillators,
the argument in this section has the merit
that it does not depend on the particular background
of the closed string. It also illuminates
the (world-sheet) geometrical nature of the relation.

We first recall the basic constraints of the consistent
boundary state in a general background.
First, it must be invariant under the
conformal transformation at the boundary:
\begin{equation}
\label{eq:weak_rel}
\mathcal{L}_n|B\rangle =0\,,\quad
\mathcal{L}_n\equiv L_n-\tilde L_{-n}\,.
\end{equation}
This linear condition alone, however, is not sufficient.
In order to have a well-defined Hilbert space in the open string
channel, we need to impose the Cardy condition \cite{r:Cardy}; that is,
for two arbitrary boundary states $|B\rangle$ and $|B'\rangle$,
we calculate the inner product with a propagator
and apply the modular transformation,
\begin{equation}
\label{eq:strong_rel}
\langle B|q^{\frac{1}{2}(L_0+\tilde L_0)} |B'\rangle =  \chi_{BB'}(q) 
 =  \sum_{i} N_{BB'}^i  \chi_i(\tilde q)\,.
\end{equation}
Here, the index $i$ in the summation labels the irreducible
representations,
and $\chi_i(\tilde q)$ is the corresponding character.
The Cardy condition is that the coefficient $N_{BB'}^i$
must be a non-negative integer for any $i$.
In the following,
we refer to the first requirement (\ref{eq:weak_rel}) as the ``weak
condition'' and 
the second (\ref{eq:strong_rel}) as the ``strong condition.''

We conjecture that the {\em strong}  condition is equivalent
to our idempotency relation, since both are quadratic
with respect to the string fields. 
We have confirmed this explicitly for flat backgrounds,
including toroidal compactification,
as we briefly explain in the discussion section.  
However, this is possible only for cases in which we have an explicit
oscillator representation, and a background-independent
proof  is yet to be completed.
Therefore, we present only the proof  of
the weak condition in this section.
The assertion that we demonstrate is
\begin{equation}
\mathcal{L}_n | B_a \rangle = 0 \quad (a=1,2) \quad
\rightarrow \quad \mathcal{L}_n | B_1 \star B_2 \rangle = 0
\end{equation}
in the LPP formulation. The last expression is equivalent to 
\begin{equation}
\label{eq:weak_rel2}
(\mathcal{L}_n\Phi)\cdot (\Phi_{B_1}\star \Phi_{B_2})=
\langle h_3[\wp\mathcal{L}_n \Phi] h_1[\mathcal{O}_1]h_2[\mathcal{O}_2]
\rangle =0
\,,\quad
(\mbox{for any }
\Phi)
\end{equation}
where $\mathcal{O}_r$ is the operator that corresponds
to the boundary state $|B_r\rangle=\mathcal{O}_r|0\rangle$,
up to the ghost sector and the $\alpha$ conservation factor.

In the following, we present the derivation 
in which $L_n$ and $\tilde{L}_n$ are 
restricted to the matter sector.  This restriction simplifies the
computation and helps to illuminate the
essence of our proof.  The ghost sector is universal for
any background, and in a previous paper \cite{r:kmw1}
we gave a proof in the operator formulation
for the case of a flat background.
For this reason, ignoring the ghost sector will not be so problematic.

The Virasoro generators $L_n$ are originally defined
as the coefficients of the stress energy tensor in $w_r$ coordinates,
$T(w_r)=\sum_n L_n w_r^{-n-2}$.  For our purpose,
it is more convenient to use the coordinates 
$\zeta_r=\log(w_r)$,
because in this case, the connection conditions
between three patches at the vertex become the simplest.
The conformal transformation of the stress-energy tensor is the standard
one,
\begin{equation}
T(\zeta) =\left(\frac{dw}{d\zeta}\right)^2 T(w) +\frac{c}{12} 
S(w,\zeta)\,,\quad
S(z,w)=\frac{\frac{d^3 z}{dw^3}}{\frac{dz}{dw}}-\frac{3}{2} \left(
\frac{\frac{d^2 z}{dw^2}}{\frac{dz}{dw}}\right)^2\,.
\end{equation}
In our case, the stress energy tensor is written
$T(\zeta)=\sum_{n}L_ne^{-n\zeta}-\frac{c}{24}$.

At the boundary of the disk $|w|=1$ [or $\mathrm{Re}(\zeta)=0$],
the operators $\mathcal{L}_n$ appear in the combination
\begin{equation}
 \mathcal{T}(\sigma)= \sum_{n
}\mathcal{L}_n e^{-in\sigma}
 = T(i\sigma) - \tilde{T}(-i\sigma)\,.
\end{equation}
This operator describes the reparametrization at the boundary.
A (boundary) primary field of dimension $\Delta$
has the following relation with $\mathcal{L}_n$:
\begin{equation}
 \mathcal{L}_n V(\sigma) |B\rangle = e^{in\sigma}\left(
-i\frac{d}{d\sigma}+n\Delta \right)V(\sigma) |B\rangle\,.
\end{equation}
This corresponds to the reparametrization $\delta \sigma =-i e^{in\sigma}$.
As a special example, the operator $\mathcal{T}(\sigma)$ itself
transforms under the boundary reparametrization according to
\begin{equation}
 \mathcal{T}(\sigma) =\mathcal{T}(\sigma')
\left(\frac{d\sigma'}{d\sigma}\right)^2\,.
\end{equation}
There is no anomaly term here, as there is no central
extension for $\mathcal{L}_n$.

In the following, we replace the expression (\ref{eq:weak_rel2})
by applying the overall conformal transformation $\rho(z)$:
\begin{equation}
\label{eq_boundary_V}
\langle f_3[\wp\mathcal{L}_n \Phi] f_1[\mathcal{O}_1]f_2[\mathcal{O}_2]
\rangle =0\,,\quad (\mbox{for any }\Phi)
\end{equation}
where 
\begin{equation}
\label{eq_boundary_V2}
f_r(w_r)=\rho(h_r(w_r))=\alpha_r \log w_r +\tau_0+i\beta_r
=\alpha_r \zeta_r +\tau_0+i\beta_r
\,.
\end{equation}
The correlation function on the left-hand side should
be evaluated on Mandelstam diagram.
The generalized gluing and resmoothing theorem (GGRT)
\cite{r:LPP2, r:SenS, r:AKT1} 
ensures that this expression is proportional to the original one,
Eq.~(\ref{eq:weak_rel2}).

With the above preparation, the proof of the assertion is straightforward.
The strategy is to rewrite the operator $\mathcal{L}_n$
in terms of the contour integration \sloppy
 $\int_{-\pi}^\pi \frac{d\sigma_3}{2\pi}
e^{in\sigma_3}\mathcal{T}_3(\sigma_3)$ and to use the connection condition
at the boundary that is implied by the 3-string vertex.
This makes these operators act on $\mathcal{O}_r$ 
and annihilate them under the assumption.
In this process, we need to change the coordinate $\sigma_3$ to 
$\sigma_r$ ($r=1,2$).  We write the corresponding boundary
stress-energy tensors as $\mathcal{T}_r(\sigma_r)$. These are
related as
$\mathcal{T}_3(\sigma_3)=\mathcal{T}_r(\sigma_r)(\alpha_3/\alpha_r)^2$,
from Eq.~(\ref{eq_boundary_V2}).
We also use the following additional notation:
\begin{eqnarray}
\label{eq:wpLn}
 &&\wp \mathcal{L}_n=\mathcal{L}_n\wp_n,\quad
\wp_n\equiv \oint {d\sigma\over 2\pi}e^{i\sigma(\mathcal{L}_0-n)},
\\
&&\Sigma_r(\sigma_r) \equiv f_3^{-1}(f_r(\sigma_r))=
\frac{1}{\alpha_3}(\alpha_r \sigma_r+\beta_r-\beta_3).
\end{eqnarray}
Then, the proof is given as follows:
\begin{eqnarray}
 && \langle f_3[\wp\mathcal{L}_n \Phi] f_1[\mathcal{O}_1]f_2[\mathcal{O}_2]
\rangle = 
\langle f_3[\mathcal{L}_n \wp_n \Phi] f_1[\mathcal{O}_1]f_2[\mathcal{O}_2]
\rangle\nonumber\\
&& = \int_{-\pi}^\pi \frac{d\sigma_3}{2\pi} e^{in\sigma_3}
\langle f_3[\mathcal{T}_3(\sigma_3) 
\wp_n \Phi] f_1[\mathcal{O}_1]f_2[\mathcal{O}_2]
\rangle \nonumber\\
&& = 
-\int_{-\pi}^\pi \frac{d\sigma_1}{2\pi}e^{in \Sigma_1(\sigma_1)}
\left(\frac{d\Sigma_1}{d\sigma_1}\right)^{-1}\langle
f_3[\wp_n \Phi] f_1[\mathcal{T}_1(\sigma_1)\mathcal{O}_1]
f_2[\mathcal{O}_2]\rangle\nonumber\\
&& \quad -\int_{0}^{2\pi} \frac{d\sigma_2}{2\pi}e^{in \Sigma_2(\sigma_2)}
\left(\frac{d\Sigma_2}{d\sigma_2}\right)^{-1}\langle
f_3[\wp_n \Phi] f_1[\mathcal{O}_1]
f_2[\mathcal{T}_2(\sigma_2)\mathcal{O}_2]\rangle =0\,.
\label{eq:proof}
\end{eqnarray}
We note that we have changed the integration range of 
$\sigma_2$
in the last expression.  This is because 
the function $\Sigma_2$ has a jump at $\sigma_2=0$,
while it is continuous at $\sigma_2=\pm \pi$, if we require
periodic boundary conditions for $\Sigma_2$. 

While our proof looks straightforward, there is a subtle point
that must be treated carefully.
This regards the singularity at the interaction point.
In the final line of Eq.~(\ref{eq:proof}), there appear functions of the
form $e^{in\Sigma_r(\sigma_r)}$.  These functions are, as we previously
mentioned, not continuous. While this does not seem problematic in the
above calculation, we have to treat it more seriously in the argument
given in the next section.

%%%%%%%%%%%%%%%%%%%%%%%%%%%%%%

\section{Derivation of the open string spectrum
\label{sec:spectrum}}

%%%%%%%%%%%%%%%%%%%%%%%%%%%%%%

\subsection{Fluctuations around the idempotency equation}

%%%%%%%%%%%%%%%%%%%%%%%%%%%%%%

In Ref.~\citen{r:kmw1}, we examined variations of 
the idempotency relation
(\ref{eq_idempotency3})
and derived the on-shell conditions for the open string modes.
The analysis was, however, restricted to lower exited states,
namely the tachyon and the massless vector particle.
Analysis for the higher excited modes is not feasible, because
the computation becomes complicated.
In the LPP approach, however, we can carry out a more systematic study.
The idempotency equation (\ref{eq_idempotency2}) can be rewritten
in LPP language as
\begin{eqnarray}
\label{eq:LPPBB=B}
 &&\langle h_1[b_0^-\Phi_B(x^{\perp})]\,h_2[b_0^-\Phi_B(y^{\perp})
]\,h_3[b_0^-\wp \,\Phi]\rangle\\
&&=-\delta^{d-p-1}(x^{\perp}-y^{\perp})
{\cal{C}} \langle I[c^+_0\Phi_B(x^{\perp})]\,b_0^-\Phi\rangle\,,
\nonumber
\end{eqnarray}
where we have assigned $\alpha=-\alpha_1-\alpha_2$ to the
arbitrary $\Phi$ and dropped the factor of $2\pi\delta(0)$
in the $\alpha$ sector.
The variation of the idempotency equation
(\ref{eq_idempotency3}) 
for the deformation Eq.~(\ref{eq_deform}) in the  matter sector 
can be rewritten in the same way.
As in the previous section, we need to apply the conformal transformations
$\rho(z)$ to the left  and $f_3$ to the right
to use the coordinate system $\zeta$.  
The GGRT implies that the relation (\ref{eq:LPPBB=B})
is unchanged as long as the total central charge vanishes, i.e.,
for the critical dimension.  
 The left-hand side of Eq.~(\ref{eq_idempotency3}) becomes
 (with $J\equiv f_3\circ I$)
\begin{eqnarray}
&& \langle f_1[b_0^-\delta \Phi_B(x^{\perp})]\,
f_2[b_0^-\Phi_B(y^{\perp})]\,f_3[b_0^-\wp \,\Phi]\rangle\nonumber\\
&&+
\langle f_1[b_0^-\Phi_B(x^{\perp})]\,
f_2[b_0^-\delta\Phi_B(y^{\perp})]\,f_3[b_0^-\wp \,\Phi]\rangle\nonumber\\
&=&\oint {d\sigma_1\over 2\pi}
\langle f_1[b_0^-V(\sigma_1)\Phi_B(x^{\perp})]\,
f_2[b_0^-\Phi_B(y^{\perp})]\,f_3[b_0^-\wp \,\Phi]\rangle\nonumber\\
&&+\oint {d\sigma_2\over 2\pi}
\langle f_1[b_0^-\Phi_B(x^{\perp})]\,
f_2[b_0^-V(\sigma_2)\Phi_B(y^{\perp})]\,f_3[b_0^-\wp \,\Phi]\rangle\nonumber\\
&=&\left\langle f_1[b_0^-\Phi_B(x^{\perp})]\,
f_2[b_0^-\Phi_B(y^{\perp})]\,f_3\biggl[\biggl(
\oint{d\sigma_1\over 2\pi}\Sigma_1[V(\sigma_1)]
+
\oint{d\sigma_2\over 2\pi}\Sigma_2[V(\sigma_2)]
\biggr)b_0^-\wp\Phi\biggr]\right\rangle\nonumber\\
&=&-\delta^{d-p-1}(x^{\perp}-y^{\perp})\,
{\cal{C}}\nonumber\\
&&~~~~~\times \left\langle J[c^+_0
 \Phi_B(x^{\perp})]\,f_3\left[b_0^-
\left(
\oint{d\sigma_1\over 2\pi}\Sigma_1[V(\sigma_1)]
+
\oint{d\sigma_2\over 2\pi}\Sigma_2[V(\sigma_2)]
\right)\wp\Phi\right]\right\rangle\nonumber\\
&=&-\delta^{d-p-1}(x^{\perp}-y^{\perp})\,
{\cal{C}}\nonumber\\
&&~~~~~\times \left\langle J\left[c^+_0b_0^-
\wp\left(
\oint{d\sigma_1\over 2\pi}\Sigma_1[V(\sigma_1)]
+
\oint{d\sigma_2\over 2\pi}\Sigma_2[V(\sigma_2)]
\right)
\Phi_B(x^{\perp})\right]
f_3[\Phi]\right\rangle,
\label{eq:LHSdelta}
\end{eqnarray}
where we have used Eq.~(\ref{eq:LPPBB=B}).
Similarly, the right-hand side of Eq.~(\ref{eq_idempotency3}) becomes
\begin{eqnarray}
 &&-\delta^{d-p-1}(x^{\perp}-y^{\perp})\,{\cal{C}}
\langle J[c^+_0\delta\Phi_B(x^{\perp})]\,f_3[b_0^-\Phi]\rangle
\nonumber\\
&&=\delta^{d-p-1}(x^{\perp}-y^{\perp})\,{\cal{C}}
\oint {d\sigma_3\over 2\pi}
\langle J[c^+_0 b_0^-V(\sigma_3)\Phi_B(x^{\perp})]\,f_3[\Phi]\rangle\,.
\label{eq:RHSdelta}
\end{eqnarray}
The two equations (\ref{eq:LHSdelta}) and (\ref{eq:RHSdelta})
imply that the variation equation (\ref{eq_idempotency3})
is reduced to
\begin{eqnarray}
\label{eq:Vcondition}
\wp\left(\oint{d\sigma_1\over 2\pi}\Sigma_1[V(\sigma_1)]
+
\oint{d\sigma_2\over 2\pi}\Sigma_2[V(\sigma_2)]
+\oint {d\sigma_3\over 2\pi}V(\sigma_3)\right)|B(x^{\perp})\rangle=0\,,
\end{eqnarray}
where we have used
$|B(x^{\perp})\rangle=c_0^+b_0^-|\Phi_B(x^{\perp})\rangle$.
This is solved by the requirement
\begin{eqnarray}
 \Sigma_r[V(\sigma_r)]\,|B(x^{\perp})\rangle
={d\over d\sigma_r}\Sigma_r(\sigma_r)\,
V(\Sigma_r(\sigma_r))\,|B(x^{\perp})\rangle\,.
\label{eq:weight1}
\end{eqnarray}
With this condition, the integrations in
Eq.~(\ref{eq:Vcondition}) cancel, because the corresponding
contours in the $z$-plane are $C_1,C_2$ and $-C_1-C_2$ in
Fig.~\ref{fig:Mandel}, respectively.

Suppose the conformal mappings $\Sigma_r$ are generic. Then,
the infinitesimal form of  Eq.~(\ref{eq:weight1}) becomes
\begin{eqnarray}
\label{eq:onshell}
 \mathcal{L}_n V(\sigma)
|B(x^{\perp})\rangle
=e^{in\sigma}\left(-i{d\over d\sigma}+n\right)V(\sigma)
|B(x^{\perp})\rangle\,.
\end{eqnarray}
This implies that $V(\sigma)$ must be a boundary 
primary operator with weight 1.
This provides a sufficient condition to solve
Eq.~(\ref{eq_idempotency3}), and
at the same time, it is identical to the physical state condition for
the open string. We find that the entire {\it open} string spectrum
is contained as a solution to
Eq.~(\ref{eq_idempotency3}), which is written in terms of the {\it
closed} string variable.
We note that the condition (\ref{eq:onshell}) implies
the BRST invariance of the deformation, i.e.,
\begin{eqnarray}
 Q_B\oint{d\sigma\over 2\pi}V(\sigma)|B(x^{\perp})\rangle=0\,,
\end{eqnarray}
where the BRST operator $Q_B$ is defined in Eq.~(\ref{eq:BRST}).
This can be proved using the identity
\begin{eqnarray}
 Q_B{\cal O}_{\rm{matter}}|B(x^{\perp})\rangle
=\sum_{n=-\infty}^{\infty}c_{-n}
\mathcal{L}_n
{\cal O}_{\rm{matter}}|B(x^{\perp})\rangle\,.
\end{eqnarray}

%%%%%%%%%%%%%%%%%%%%%%%%%%%%%%%%
\subsection{Constraint from the interaction point}
%%%%%%%%%%%%%%%%%%%%%%%%%%%%%%%%
Actually, the mappings $\Sigma_r$ are {\em not} generic but, rather, 
linear transformations with a  discontinuity at the interaction point.  
In this sense, it is not obvious whether or not all the solutions to 
Eq.~(\ref{eq:Vcondition}) are  restricted to the boundary primary fields
of dimension 1.
Indeed, if we ignore the subtlety at the interaction point,
the constraint from Eq.~(\ref{eq:weight1}) does not imply 
that the operator $V(\sigma)$ must be {\em primary},
since the conformal transformation is restricted to linear
transformations. 
In this sense, additional nontrivial constraints should come only from the
interaction point.  Instead of treating the generic vertex
$V(\sigma)$, which would
be technically difficult, we examine the lower excited mode explicitly.
In particular, we pick the example of the massless vector particle.
This is interesting, because it is difficult to confirm the transversality
condition in the operator formalism,\cite{r:kmw1}
due to the infinite-dimensionality of the Neumann coefficients.
The computation involves a derivation of the {\em large} 
conformal transformation  for the non-primary fields.

We first calculate the transformations of the
scalar-type and vector-type operators $V_S$ and $V_V$,
which we considered in Ref.~\citen{r:kmw1}, and the modified 
version $\hat{V}_V$ \cite{r:MN}:
\begin{eqnarray}
 &&V_S(\sigma)=\,:e^{ik_{\mu}X^{\mu}(\sigma)}:\,,~~
V_V(\sigma)=\,:\zeta_{\mu}\partial_{\sigma}X^{\mu}(\sigma)
e^{ik_{\nu}X^{\nu}(\sigma)}:\,,\\
&&\hat{V}_{V}(\sigma)\equiv V_V(\sigma)
-(\zeta_{\mu}\theta^{\mu\nu}k_{\nu}/4\pi)V_S(\sigma)\,.
\end{eqnarray}
Here,
$\theta\equiv \pi({\cal{O}}-{\cal{O}}^{T})/2=-2\pi(1+F)^{-1}F(1-F)^{-1}$
 is the noncommutativity parameter.\cite{r:SW}
For these operators, after a straightforward computation \cite{r:MN},
we obtain
\begin{eqnarray}
\mathcal{L}_n V_S(\sigma)|B(x^{\perp})\rangle
&=&e^{in\sigma}\left(-i\partial_{\sigma}+n
\Delta\right)
V_S(\sigma)|B(x^{\perp})\rangle,\\
\mathcal{L}_n \hat{V}_V(\sigma)|B(x^{\perp})\rangle
&=&e^{in\sigma}\left(-i\partial_{\sigma}
+n(\Delta+1)\right)
\hat{V}_V(\sigma)|B(x^{\perp})\rangle\nonumber\\
&&
+e^{in\sigma}n^2\, i\,\Xi\, V_S(\sigma)
|B(x^{\perp})\rangle,~~~~
\end{eqnarray}
where we have defined
$\Delta\equiv k_{\mu}G^{\mu\nu}k_{\nu}/2$ and $\Xi\equiv -i\zeta_{\mu}
G^{\mu\nu}k_{\nu}/2$.
The open string metric $G^{\mu\nu}$ appears due to the boundary state
$|B(x^{\perp})\rangle$ given in Eq.~(\ref{eq_boundary_state}).
These relations imply %the infinitesimal transformation,
\begin{eqnarray}
 \delta_{\epsilon}V_S(\sigma)|B(x^{\perp})\rangle
&=&\epsilon(\sigma)\partial_{\sigma}V_S(\sigma)|B(x^{\perp})\rangle
+\Delta\partial_{\sigma}\epsilon(\sigma)V_S(\sigma)|B(x^{\perp})\rangle\,,\\
 \delta_{\epsilon}\hat{V}_V(\sigma)|B(x^{\perp})\rangle&=&
\epsilon(\sigma)\partial_{\sigma}\hat{V}_V(\sigma)|B(x^{\perp})\rangle
+(\Delta+1)\partial_{\sigma}\epsilon(\sigma)
\hat{V}_V(\sigma)|B(x^{\perp})\rangle\nonumber\\
&&+\,\Xi\,\partial^2_{\sigma}\epsilon(\sigma)V_S(\sigma)|B(x^{\perp})\rangle
\end{eqnarray}
for the infinitesimal transformation
$\delta_{\epsilon}\sigma=-\sum_{n}\epsilon_n ie^{in\sigma}
=\epsilon(\sigma)$.
The last term in the second equation shows that 
 $\hat{V}_V$ is not primary unless $\Xi=0$.
A finite transformation should be determined from
this infinitesimal deformation by the requirement of the
cocycle condition, which states that under a sequence of coordinate
transformations
$\sigma_1\rightarrow \sigma_2 \rightarrow \sigma_3$, the combination
of two transformations $\sigma_1\rightarrow \sigma_2$ followed by
$\sigma_2\rightarrow \sigma_3$
is identical to the direct one, $\sigma_1\rightarrow \sigma_3$.
The transformations of $V_S$ and $\hat{V}_T$ that satisfy this condition
are
\begin{eqnarray}
 (d\sigma)^{\Delta}\,V_S(\sigma)|B(x^{\perp})\rangle
&=&(d\lambda)^{\Delta}\,
V_S(\lambda)|B(x^{\perp})\rangle\,,\\
(d\sigma)^{\Delta+1} \hat{V}_V(\sigma)|B(x^{\perp})\rangle
&=&(d\lambda)^{\Delta+1}\left(\hat{V}_V(\lambda)|B(x^{\perp})\rangle
-\Xi\, {\partial_{\lambda}^2\sigma\over \partial_{\lambda}\sigma}V_S(\lambda)
|B(x^{\perp})\rangle\right).~~~~~
\label{eq:hatVVtrans}
\end{eqnarray}

The relevant formula is obtained by setting
$\sigma=\sigma_r,\lambda=\sigma_3=\Sigma_r(\sigma_r)$ (for $r=1,2$).
Except at the interaction point, the anomalous term vanishes,
since $\Sigma_r$ is a linear function.
Therefore, the only constraint for $V=\hat{V}_V$
is the on-shell condition, $\Delta=k_{\mu}G^{\mu\nu}k_{\nu}/2=0$.

\begin{figure}[htbp]
          \begin{center}
          \scalebox{0.8}[0.8]{\includegraphics{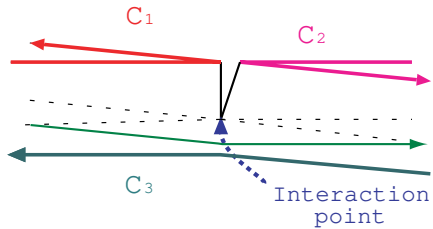}}
          \end{center}
          \caption{Landscape near the interaction point.}
          \label{fig:interaction_pt3}
\end{figure}

At the interaction point, the anomaly term diverges, due to
the discontinuity of $\Sigma_r$. Geometrically, this results from
the  curvature singularity.
In Fig.~\ref{fig:interaction_pt3}, we plot the landscape 
of the world-sheet at the interaction point.
The three paths $C_r$ are the contours of $\sigma_r$ in
Eq.~(\ref{eq:Vcondition}),
shifted slightly from the singularity. We use this shift to define the
regularization.

\begin{figure}[htbp]
          \begin{center}
          \scalebox{0.8}[0.8]{\includegraphics{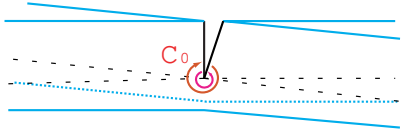}}
          \end{center}
          \caption{The contour $C_1+C_2+C_3$ 
in Fig. \ref{fig:interaction_pt3}
can be deformed into $C_0$ around the interaction 
point for $\hat{V}_V$ with $\Delta=0$.}
          \label{fig:interaction_pt4}
\end{figure}

If we impose the condition $\Delta=0$, the contribution from 
most of the contour of integration mutually cancels.
The only remaining part is the integration 
around the interaction point (see Fig.~\ref{fig:interaction_pt4}).
At the interaction point, $\rho$ does not give a proper parametrization,
since $\arg(\rho)$ increases by $4\pi$ along the contour $C_0$.
It is therefore necessary  to make a (large) transformation from $\rho$ to
$z$ in the $z$-plane (see Fig.~\ref{fig:Mandel}).
We use $\sigma={\rho(z)/(i\alpha_3)}+{\rm const.}$ [which corresponds to
$\sigma_3$, according to Eq.~(\ref{eq:hrdef})]
and $\lambda=z$ in Eq.~(\ref{eq:hatVVtrans}) to evaluate the integration
in Eq.~(\ref{eq:Vcondition}). 
Noting that $d\rho/dz\sim {\rm const.}(z-z_0)$ near 
the interaction point $z_0$,
we have
\begin{eqnarray}
&& d\sigma\hat{V}_V(\sigma)|B(x^{\perp})\rangle\nonumber\\
&&=dz \left[\hat{V}_V(z)|B(x^{\perp})\rangle
-\Xi\, \left((z-z_0)^{-1}+{\cal O}((z-z_0)^0)\right) V_S(z)
|B(x^{\perp})\rangle\right]
\end{eqnarray}
for $\Delta=0$. 
Note that the anomalous term gives a pole residue.
Therefore, the left-hand side of Eq.~(\ref{eq:Vcondition}) for 
$\hat{V}_V(\sigma)$ is evaluated as
\begin{eqnarray}
 i \wp\,\Xi\, V_S(z_0)|B(x^{\perp})\rangle
=i\,\Xi\oint{d\sigma\over 2\pi}V_S(\sigma)|B(x^{\perp})\rangle
\end{eqnarray}
for $\Delta=0$. We conclude that 
not only the mass-shell condition $k_{\mu} G^{\mu\nu}k_{\nu}=0$ 
but also the transversality condition
$\Xi=\zeta_{\mu}G^{\mu\nu}k_{\nu}/2=0$ should be imposed
for the solution 
$\delta\Phi_B=\oint{d\sigma\over 2\pi}\hat{V}_V(\sigma)\Phi_B$ 
to Eq.~(\ref{eq_idempotency3}).

%%%%%%%%%%%%%%%%%%%%%%%%%%%%%%%%%%%%%%%%%%%%%%%%%%%%%%%%%%%%%%
\section{Summary and future problems}
\label{sec_conjecture}
%%%%%%%%%%%%%%%%%%%%%%%%%%%%%%%%%%%%%%%%%%%%%%%%%%%%%%%%%%%%%%
In this paper, we have derived some results
which supplement our previous paper \cite{r:kmw1} with regard to the
operator formulation.  They include the determination
of the analytic form of the coefficient of the idempotency
relation and a generalization to the case of the Witten-type
closed string vertex.  The former implies that
a large class of boundary states in the flat
background satisfy a universal nonlinear relation.

We also used the representation of the HIKKO vertex
in the CFT language and examined the idempotency relation
in a background-independent fashion.
There are some additional merits of
this new computation.  For example, we can show
that the transversality condition of
the massless vector particle
is necessary.  In the operator formalism,
this is difficult, due to the appearance of
the divergence.

We note that our proof in the case of a generic background
is restricted to a confirmation of
the weak condition.  While this implies that the
product of the boundary states is a linear combination
of the boundary states, it does not fix the coefficients.
In order to determine them, the Cardy condition is needed.

Because both the Cardy condition and the idempotency relation
are quadratic in the string field, it is natural to
conjecture that these two conditions are equivalent.
For the flat background, this is indeed the case.
The generic boundary states that satisfy the condition
(\ref{eq:weak_rel})
are called Ishibashi states.\cite{r:Ishibashi} For the flat background,
they are given by the Fourier transformation
of (\ref{eq_boundary_state}) with respect to $x^\perp$,
and we denote them as $|p^\perp\rangle\!\rangle$.
After the inclusion of the $\alpha$ parameter dependence
and the insertion of the ghost zero modes, the star product
of Ishibashi states is written
\begin{equation}
\label{eq:p*p=p+p}
   |p^\perp,\alpha_1\rangle\!\rangle
   \star |q^\perp,\alpha_2\rangle\!\rangle
   = \mathcal{C}c_0^+
|p^\perp+q^\perp,\alpha_1+\alpha_2\rangle\!\rangle\,.
\end{equation}
We note that the star product is not diagonal in this basis.
It takes the form of the idempotency relation only when
we take a linear combination (Fourier transformation)
of Ishibashi states to form the Cardy states.

We conjecture that a relation similar to Eq.~(\ref{eq:p*p=p+p}) holds
for more general backgrounds.  We write the generic Ishibashi state as
$|i\rangle\!\rangle$, where $i$ is the label of the highest
weight representation.  Then, a possible generalization of
the above relation is
\begin{equation}
   |i\rangle\!\rangle \star |j\rangle\!\rangle \propto
   \sum_{k} {N_{ij}}^k |k\rangle\!\rangle\,,
\end{equation}
where ${N_{ij}}^k$ is Verlinde's fusion coefficient.
By taking linear combinations of the form
$|a\rangle=\sum_i\frac{S_{ia}}{\sqrt{S_{i0}}}|i\rangle\!\rangle$,
where $S_{ia}$ is the modular transformation matrix, the above
relation is diagonalized with respect to the Cardy states $|a\rangle$:
\begin{equation}
   |a\rangle\star|b\rangle \propto \delta_{ab}|b\rangle\,.
\end{equation}
In the future, we will report our computation in the case of
more generic backgrounds,
including toroidal and orbifold compactification,
to examine this conjecture.

As another important issue,
we have observed that a VSFT-like scenario seems possible
with the closed string variables.
While this seems more hypothetical,
there are some points that must be clarified in order to study
this scenario more deeply. 
We have already commented that there seem to be no closed string
physical states, even if we expand the action (\ref{eq:action_general})
around the boundary state.
If this is true, our Lagrangian would seem to
describe only open strings, while we have used closed string
variables. In order to ascertain the consistency of this approach, the most
nontrivial test would be to calculate the perturbative amplitudes such
as the Veneziano amplitude.  At that stage, we will need to clarify
the nature of the $\alpha$ parameter.
A natural interpretation may be that it is related to the
moduli space of an open Riemann surface.

There are also various interesting directions in which
our computation can be generalized.  These include (i) a
computation on a nontrivial background, which we have
already mentioned,
(ii) derivation of the noncommutative geometry from the closed string field
theory, as we suggested in Footnote \ref{fn:Moyal},
(iii) a more detailed exploration of the properties of boundary
states with respect to the star product of the Witten-type
closed string field theory, in particular including the
higher-order interactions, and finally (iv) the supersymmetric
extension of the analysis, where the inclusion of the Ramond-Ramond
sector is a major challenge in string field theory.
The identity satisfied by the boundary states may provide information
for the solution to this important problem.

\section*{Acknowledgements}

We would like to thank K.~Hashimoto, H.~Hata and K.~Ohmori
 for valuable discussions and comments.
I.~K. would like to thank T.~Asakawa, T.~Kawano,
S.~Matsuura and T.~Takahashi
for useful conversations.
I.~K. is supported in part by JSPS Research Fellowships
for Young Scientists. Y.~M. is supported in part by Grant-in-Aid (\#
13640267) from the Ministry of Education, Science, Sports and Culture of
Japan.

\appendix

%%%%%%%%%%%%%%%%%%%%%%%%%%%%%%%%%%%%%%%%%%%%%%%%%%%%%%%%%%%%%%
\section{Notation and Conventions
\label{sec:osc-LPP}}
%%%%%%%%%%%%%%%%%%%%%%%%%%%%%%%%%%%%%%%%%%%%%%%%%%%%%%%%%%%%%%

In this section, we explicitly demonstrate the correspondence
between the notation of HIKKO and of LPP.
This is convenient for the purpose of comparing the results 
obtained in oscillator language with the original HIKKO notation
(for example our previous paper \cite{r:kmw1})
 and those obtained in the LPP language.

First, the HIKKO oscillators
$\alpha^{(+)}_n,\alpha^{(-)}_n,c_n^{(+)},c_n^{(-)},\bar{c}_n^{(+)},
\bar{c}_n^{(-)}$ 
correspond to the LPP oscillators (or in the conventional CFT notation)
$\alpha_n,\tilde{\alpha}_n,c_n,\tilde{c}_n,b_n,\tilde{b}_n$,
respectively.
In particular, the ghost zero modes are related as
\begin{eqnarray}
\left({\partial\over \partial \bar{c}_0}
={1\over 2}(c^{(+)}_0+c^{(-)}_0)\right)_{\rm H}
&=&\left(c^+_0= {1\over 2}(c_0+\tilde{c}_0)\right)_{\rm L},\nonumber\\
\left(i{\partial\over \partial \pi_c^0}=c^{(+)}_0-c^{(-)}_0\right)_{\rm H}
&=&\left(c^-_0=c_0-\tilde{c}_0\right)_{\rm L},\nonumber\\
\left(\bar{c}_0=\bar{c}^{(+)}_0+\bar{c}^{(-)}_0\right)_{\rm H}
&=&\left(b^+_0=b_0+\tilde{b}_0\right)_{\rm L},\nonumber\\
\left(-i\pi_c^0={1\over 2}(\bar{c}_0^{(+)}-\bar{c}_0^{(-)})\right)_{\rm H}
&=&\left(b^-_0={1\over 2}(b_0-\tilde{b}_0)\right)_{\rm L}.~~~~~
\label{eq:gh_zeromodes}
\end{eqnarray}
We often indicate a quantity in the original HIKKO notation by the
subscript H and a quantity in the LPP notation by the subscript L in
this section. (In other sections, we have used the
LPP notation and did not include the subscript L.)

In bra-ket notation, we identify the vacuum convention by
\begin{eqnarray}
\label{eq:vac_HL}
{}_{\rm H}\langle 0|=
{}_{\rm L}\langle 1,1|c_0^-c_0^+\,,~~~~~~~
|0\rangle_{\rm H} = | 1,1\rangle_{\rm L}\,,
\end{eqnarray}
where
$ {}_{\rm L}\langle 1,1|:=\langle 0|c_{-1}\tilde{c}_{-1},~~
| 1,1\rangle_{\rm L}=c_1\tilde{c}_1|0\rangle$,
and $\langle 0|$ and $|0\rangle$ denote conformal vacua and are
normalized as
$\langle 1,1|c_0\tilde{c}_0|1,1\rangle
=\langle 1,1|c^-_0c^+_0|1,1\rangle
=(2\pi)^d\delta^d(0)=V_d$.
The string field in the LPP formulation can be obtained 
from that in the HIKKO formulation by Fourier transformation
with respect to ghost zero mode:
\begin{eqnarray}
 \label{eq:map}
&&-i\int d\bar{c}_0 d\pi_c^0 e^{-\bar{c}_0c^+_0+i\pi_c^0c_0^-}
|\Phi(\alpha)\rangle_{\rm{H}}
= |\Phi(\alpha)\rangle_{\rm{L}}\,.
\end{eqnarray}
This implies that the ghost zero mode expansion becomes
\begin{eqnarray}
&&-i\int d\bar{c}_0 d\pi_c^0 e^{ 
-\bar{c}_0c^+_0+i\pi_c^0c_0^-}
\left[-{\bar{c}_0|\phi\rangle+|\psi\rangle
+\bar{c}_0\pi_c^0|\chi\rangle+i\pi_c^0|\eta\rangle}\right]_{\rm H}
\nonumber\\
&&= \left[
c_0^-|\phi\rangle+c_0^-c_0^+|\psi\rangle
+i|\chi\rangle-c_0^+|\eta\rangle\right]_{\rm L}.
\end{eqnarray}
As expected, the physical sector $\bar{c}_0|\phi\rangle$
in the HIKKO formulation given in Ref.~\citen{r:HIKKO2} is mapped 
to $c_0^-|\phi\rangle$ in the LPP convention.

\subsection{Reflector}

The reflector in the HIKKO theory  \cite{r:HIKKO2} is given by
\begin{eqnarray}
\label{eq:HIKKO_R}
\langle \tilde{R}(1,2)|&=&
\int {d^dp_1\over (2\pi)^d}{d^dp_2\over (2\pi)^d}
{d\alpha_1\over 2\pi}{d\alpha_2\over 2\pi} \langle p_1,\alpha_1|
\langle p_2,\alpha_2|\\
&&\times\,e^{-\sum_{\pm,n\ge 1}(
a_n^{(\pm)(1)}a_n^{(\pm)(2)}
+c_n^{(\pm)(1)}\bar{c}_n^{(\pm)(2)}
-\bar{c}_n^{(\pm)(1)}c_n^{(\pm)(2)})}
\nonumber\\
&&\times (2\pi)^d\delta^d(p_1+p_2)\delta(\pi_c^{0(1)}-\pi_c^{0(2)})
\delta(\bar{c}_0^{(1)}-\bar{c}_0^{(2)})
2\pi\delta(\alpha_1+\alpha_2)\,,\nonumber
\end{eqnarray}
where we have used the momentum representation for
the matter zero mode.
We adopted the bra-ket notation for the 
$p_{\mu},\alpha$ part, in addition to the nonzero mode oscillator sector.
We can obtain the bra from the ket using the reflector
(\ref{eq:HIKKO_R}):
${}_2\langle\Phi(-\alpha)|=\int d\bar{c}_0^{(1)} d\pi_c^{0(1)}
\langle \tilde{R}(1,2)|\Phi(\alpha)\rangle_1$.
In particular, we need to carry out the zero modes integration
in this convention.
The dot product of string fields is defined 
with the ``metric'' $\pi_c^0$ :
\begin{eqnarray}
 \Phi_1\cdot \Phi_2&=&\int d\bar{c}_0^{(1)}d\pi_c^{0(1)}
d\bar{c}_0^{(2)} d\pi_c^{0(2)}
\langle\tilde{R}(2,1)|\pi^{0(1)}_c
|\Phi_1\rangle_1|\Phi_2\rangle_2\nonumber\\
&=&(-1)^{|\Phi_1||\Phi_2|}\Phi_2\cdot \Phi_1\,.
\end{eqnarray}

Next, we review the LPP reflector and relate it to the HIKKO reflector.
It is a kind of two-string vertex and therefore is determined by 
fixing two conformal mappings.
We take $h_1(z)=I(z):=1/z$ and $h_2(z)=z$ as conformal mappings:
\begin{eqnarray}
\label{eq:refl_LPP}
 \langle R(1,2)|
&=&\int {d^dp_1\over (2\pi)^d}{d^dp_2\over (2\pi)^d}\,
\langle p_1,1,1|
\langle p_2,1,1|\\
&&\times\,e^{-\sum_{n\ge 1}(
a_n^{(1)}a_n^{(2)}
+c_n^{(1)}b_n^{(2)}
-b_n^{(1)}c_n^{(2)}+
\tilde{a}_n^{(1)}\tilde{a}_n^{(2)}
+\tilde{c}_n^{(1)}\tilde{b}_n^{(2)}
-\tilde{b}_n^{(1)}\tilde{c}_n^{(2)})
}
\nonumber\\
&&\times (2\pi)^d\delta^d(p_1+p_2)(c_0^{-(1)}+c_0^{-(2)})
(c_0^{+(1)}+c_0^{+(2)})\,.\nonumber
\end{eqnarray}
Here, we have slightly rewritten
the ghost zero mode part given in Ref.~\citen{r:LPP}
using the relation
$\langle 3,3|=\langle 1,1|c_0\tilde{c}_0 c_1\tilde{c}_1$.
The BPZ conjugate is obtained using this reflector,
$ {}_2\langle\Phi|=\langle R(1,2)|\Phi\rangle_1$,
and the BPZ inner product becomes
$\langle \Phi_1,\Phi_2\rangle=\langle R(1,2)|\Phi_1\rangle_1|\Phi_2\rangle_2
=\langle I[\Phi_1]\,\Phi_2\rangle$.
We can rewrite the inner product of the LPP string fields 
$\Phi_{1\rm L}$ and $\Phi_{2\rm L}$ with the ``metric'' $b_0^-$ 
in terms of that in the HIKKO formulation using the correspondence 
given by Eqs.~(\ref{eq:vac_HL}) and (\ref{eq:map}) as
\begin{eqnarray}
-\langle \Phi_1, b_0^-\Phi_2\rangle
&=&-\langle R(1,2)|\Phi_1\rangle_{1\rm{L}}\,b^{-(2)}_0
|\Phi_2\rangle_{2\rm{L}}\nonumber\\
&=&\int d\bar{c}_0^{(1)} d\pi_c^{0(1)}d\bar{c}_0^{(2)} d\pi_c^{0(2)}
\int {d^dp_1\over (2\pi)^d}{d^dp_2\over (2\pi)^d}\nonumber\\
&&\times\,{}_{1{\rm H}}\langle p_1|{}_{2{\rm H}}\langle p_2|
(\pi_c^{0(1)} - \pi_c^{0(2)})(\bar{c}_0^{(1)}-\bar{c}_0^{(2)})\nonumber\\
&&\times\,(2\pi)^d \delta^d(p_1+p_2)|\Phi_1\rangle_{1\rm{H}}\,\pi_c^{0(2)}
|\Phi_2\rangle_{2\rm{H}}\,.
\end{eqnarray}
Thus, including the $\alpha$ sector,
we have obtained the dot product formula as given by
Eq.~(\ref{eq:dot_Phi12}).

Finally, we note that in this paper, we include $\pi_c^0$;
that is, we do not use the so-called
$\pi_c^0$-omitted formulation, as in Refs.~\citen{r:kmw1} and \citen{r:HH}.

\subsection{3-string vertex and Neumann coefficients}

The HIKKO  3-string vertex \cite{r:HIKKO1, r:HIKKO2}
 and the LPP 3-string vertex \cite{r:LPP} are equivalent
under the above correspondence.
We will briefly demonstrate this fact. 
(See Refs.~\citen{r:KT} and \citen{r:Imai} for details.)

First, we rewrite the 3-string vertex given as a ket 
\cite{r:HIKKO2} into the form of a bra: \\
$\langle V(1,2,3)|\\:=\int d\bar{c}_0^{(1')}d\pi_c^{0(1')}
d\bar{c}_0^{(2')}d\pi_c^{0(2')}d\bar{c}_0^{(3')}d\pi_c^{0(3')}
\langle\tilde{R}(1,1')|\langle\tilde{R}(2,2')|
\langle\tilde{R}(3,3')|V(1',2',3')\rangle$.
Then, multiplying by $\pi_c^{0(3)}$ from the right
and noting the identity \\
${\alpha_1\alpha_2\over \alpha_3}\Pi_c\,\delta
\left(\sum_{r=1}^3\alpha_r^{-1}\pi^{0(r)}_c\right)\pi_c^{0(3)}
=\pi_c^{0(1)}\pi_c^{0(2)}\pi_c^{0(3)}$, with
$\alpha_1+\alpha_2+\alpha_3=0$,
we have
\begin{eqnarray}
\langle V(1,2,3)|\pi_c^{0(3)}
&=&\int\delta(1,2,3)[\mu(1,2,3)]^2
\langle p_1,\alpha_1|\langle p_2,\alpha_2|
\langle p_3,\alpha_3|
\,e^{E_{\rm{H}}(1,2,3)}
\nonumber\\
&&\times\,
\pi_c^{0(1)}\pi_c^{0(2)}\pi_c^{0(3)}
\wp^{(1)}\wp^{(2)}\wp^{(3)}\,,
\label{eq:HIKKOv3_2}\\
E_{\rm{H}}(1,2,3)&=&{1\over 2}\sum_{\pm,r,s}\sum_{n,m=0}^{\infty}
\alpha_n^{(\pm)(r)}\bar{N}^{rs}_{nm}\alpha_m^{(\pm)(s)}
-\sum_{\pm,r,s}\sum_{n,m=1}^{\infty}
c^{(\pm)(r)}_n n{\alpha_r\over \alpha_s}\bar{N}^{rs}_{nm}
\bar{c}_m^{(\pm)(s)}\nonumber\\
&&
-{1\over 2}\sum_{\pm,r,s}\sum_{n=1}^{\infty}c^{(\pm)(r)}_n (\alpha_r w^{sr}_n)
\bar{c}_0^{(s)}\,,\\
\mu(1,2,3)&=&e^{-\tau_0\sum_{r=1}^3\alpha_r^{-1}}\,,
~~~\tau_0=\sum_{r=1}^3\alpha_r\log|\alpha_r|\,,
\label{eq:mu123}
\\
\int\delta(1,2,3)&=&\int {d^dp_1\over (2\pi)^d}{d^dp_2\over (2\pi)^d}
{d^dp_3\over (2\pi)^d}
{d\alpha_1\over 2\pi}{d\alpha_2\over 2\pi}{d\alpha_3\over 2\pi}\nonumber\\
&&\times 
(2\pi)^d\delta^d(p_1+p_2+p_3)2\pi\delta(\alpha_1+\alpha_2+\alpha_3),~~~~~\\
w^{rs}_m  &=& {\alpha_r}^{-1}\left[
\delta_{r,s} \cos m\sigma_I^{(r)}
-m\sum_{n=1}^\infty \bar{N}_{mn}^{sr}\cos n\sigma_I^{(r)}
\right].
\end{eqnarray}
Here, $\sigma^{(r)}_I$ is the interaction point of the string $r$,
and $\wp^{(r)}$ is the projection operator for the level matching condition of
this string, given in Eq.~({\ref{eq:levelmatch}).
The Neumann coefficients are given by\footnote{
We can choose the three real parameters $Z_1,Z_2$ and $Z_3$ arbitrarily
as long as $p_r$ and $\alpha_r$ are conserved.
They are often chosen as $Z_1=1,Z_2=0$ and $Z_3=\infty$ [with a constant
shift in $\rho(z)$] for convenience.\cite{r:HIKKO1}
Though the Neumann coefficients in the anti-holomorphic sector are
the complex conjugates of those in the holomorphic sector in general,
those of the 3-string vertex for both the light-cone and the Witten
type are real.
}
\begin{eqnarray}
\bar{N}^{rs}_{nm}
&=&{1\over nm}\oint_0{dw_r\over 2\pi i}
\oint_0{dw_s\over 2\pi i}h'_r(w_r)h'_s(w_s){w_r^{-n}w_s^{-m}\over 
(h_r(w_r)-h_s(w_s))^2}\,,~(n,m\ge 1)~~~
\\
\bar{N}^{rs}_{n0}&=&\bar{N}^{sr}_{0n}
={1\over n}\oint_0{dw_r\over 2\pi i}h'_r(w_r){w_r^{-n}\over h_r(w_r)-h_s(0)}
\,,~~~(n\ge 1)\\
\bar{N}^{rs}_{00}&=&\log|Z_r-Z_s|=\log|h_r(0)-h_s(0)|\,,~~~(r\ne s)\\
\bar{N}^{rr}_{00}&=&-\sum_{i\ne r}{\alpha_i\over \alpha_r}\log|Z_r-Z_s|
+{1\over \alpha_r}\tau_0(Z_1,Z_2,Z_3)=\log|h'_r(0)|\,,
\end{eqnarray}
where $z=h_r(w_r)$ (with $|w_r|\le 1$), which is
a gluing map of the string $r$ into the $z$-plane, is defined by 
$\rho(z)=\alpha_r\log w_r+\tau_0(Z_1,Z_2,Z_3)
+i\pi\sum_{s=1}^{r-1}\alpha_s$, with the Mandelstam mapping
$\rho(z)=\sum_{r=1}^3\alpha_r\log(z-Z_r)$.
Here, $Z_r=h_r(0)$ and $\tau_0(Z_1,Z_2,Z_3)=\textrm{Re}\,\rho(z_0)$,
where $z_0$ is the interaction point, where we have $\rho'(z_0)=0$.
We note that the Neumann coefficients of the ghost sector
are related to those of Kunitomo-Suehiro vertex
\cite{r:Kunitomo-Suehiro} in the $P=1$ picture as
$-n(\alpha_r/\alpha_s)\bar{N}^{rs}_{nm}=N^{(1)rs}_{~~nm},\,
-{\alpha_r}w^{sr}_n=N^{(1)rs}_{~~n0}\,\,(n,m\ge 1)$.

Next, we rewrite the LPP 3-string vertex for 
the conformal mappings $h_r(w_r)\,(r=1,2,3)$, which is given by 
Eq.~(\ref{eq:LPPvertex}).
In explicit form using oscillators,
the Neumann coefficients in the matter sector
are the same as those of the HIKKO, $\bar{N}^{rs}_{nm}$.
The Neumann coefficients in the ghost sector are given by \cite{r:LPP}
\begin{eqnarray}
 N^{(gh)rs}_{~~nm}&=&\oint_0{dw_r\over 2\pi i}
\oint_0{dw_s\over 2\pi i}
(h_r'(w_r))^2(h_s'(w_s))^{-1}
{-w_r^{-n+1}w_s^{-m-2}\over h_r(w_r)-h_s(w_s)}\,,\\&&
~~~~(n\ge 2,m\ge -1)
\nonumber
\\
M^r_{in}&=&\oint_0{dw_r\over 2\pi i}(h'_r(w_r))^{-1}w_r^{-n-2}
(h_r(w_r))^{i+1}.~~(n\ge -1,\,i=-1,0,1)
\end{eqnarray}
They satisfy the following identities:
\begin{eqnarray}
&&M^r_{i,-1}=Z_r^{i+1}e^{-\bar{N}^{rr}_{00}},\\
&&\det_{r,i} M^r_{i,-1}=|Z_1-Z_2||Z_2-Z_3||Z_3-Z_1|
e^{-\sum_{r=1}^3\bar{N}^{rr}_{00}}=\mu(1,2,3)\,,~~~\\
&&N^{(1)rs}_{~~nm}=N^{(gh)rs}_{~~nm}-
\sum_{t,i}N^{(gh)rt}_{~~n,-1}
((M_{~,-1})^{-1})_{ti}M^{s}_{im}\,,~~~~(n\ge 2,m\ge 0)\\
&&N^{(1)rs}_{~~1m}=\sum_i((M_{~,-1})^{-1})_{ri}M^{s}_{im}\,,
~~~~~(m\ge 0)\\
&&{}_1\langle 3,3|{}_2\langle 3,3|{}_3\langle 3,3|\prod_{i={-1}}^1
\left(\sum_r\sum_{m\ge -1}M^r_{im}b_m^{(r)}\right)
\prod_{i={-1}}^1
\left(\sum_r\sum_{m\ge -1}M^r_{im}\tilde{b}_m^{(r)}\right)\\
&&={}_1\langle 1,1|c_0^{-(1)}c_0^{+(1)}
{}_2\langle 1,1|c_0^{-(2)}c_0^{+(2)}
{}_3\langle 1,1|c_0^{-(3)}c_0^{+(3)}\nonumber\\
&&~~~~\times\, e^{\sum_{r,s}\sum_{m\ge 0}
(c^{(r)}_1N^{(1)rs}_{~~1m}b^{(s)}_m
+\tilde{c}^{(r)}_1N^{(1)rs}_{~~1m}\tilde{b}^{(s)}_m)}
\left[\mu(1,2,3)\right]^2.\nonumber
\end{eqnarray}
Using these relations, we obtain the LPP vertex as follows:
\begin{eqnarray}
 \langle v_3|&=&\int 
{d^dp_1\over (2\pi)^d}{d^dp_2\over (2\pi)^d}
{d^dp_3\over (2\pi)^d}\nonumber\\
&&\times\,{}_1\langle p_1,1,1|c_0^{-(1)}c_0^{+(1)}
{}_2\langle p_2,1,1|c_0^{-(2)}c_0^{+(2)}
{}_3\langle p_3,1,1|c_0^{-(3)}c_0^{+(3)}\nonumber\\
&&\times(2\pi)^d\delta^d(p_1+p_2+p_3)
\,e^{E_{\rm{KS}}(1,2,3)}\left[\mu(1,2,3)\right]^2\,,
\label{eq:LPPv3_2}
\\
E_{\rm{KS}}(1,2,3)&=&{1\over 2}\sum_{r,s}\sum_{n,m=0}^{\infty}
\alpha_n^{(r)}\bar{N}^{rs}_{nm}\alpha_m^{(s)}
+{1\over 2}\sum_{r,s}\sum_{n,m=0}^{\infty}
\tilde{\alpha}_n^{(r)}\bar{N}^{rs}_{nm}\tilde{\alpha}_m^{(s)}
\nonumber\\
&&+\sum_{r,s}\sum_{n\ge 1,m\ge 0}
c^{(r)}_n N^{(1)rs}_{~~nm}
b_m^{(s)}
+\sum_{r,s}\sum_{n\ge 1,m\ge 0}
\tilde{c}^{(r)}_n N^{(1)rs}_{~~nm}
\tilde{b}_m^{(s)}\,.
\end{eqnarray}
Finally, noting the relations 
(\ref{eq:map}),~(\ref{eq:HIKKOv3_2}),~(\ref{eq:LPPv3_2})
and \sloppy
$\langle v_3|\,e^{\sum_rc_0^{+(r)}\bar{c}_0^{(r)}}
=\langle v_3|_{b_0^{(r)}\!,\tilde{b}_0^{(r)}
\rightarrow {1\over 2}\bar{c}_0^{(r)}}$,
 we have obtained the correspondence between the HIKKO
$\star$ product and the LPP 3-string vertex for the string fields,
$|\Phi_r(\alpha_r)\rangle=|\Phi_r\rangle\otimes|\alpha_r\rangle$:
\begin{eqnarray}
&&(\Phi_1(\alpha_1)\star\Phi_2(\alpha_2))\cdot\Phi_3(\alpha_3)
\nonumber\\
&=&\int d\bar{c}_0^{(1)}d\pi_c^{0(1)}d\bar{c}_0^{(2)}d\pi_c^{0(2)}
d\bar{c}_0^{(3)}d\pi_c^{0(3)}\langle V(1,2,3)|\pi_c^{0(3)}
|\Phi_1(\alpha_1)\rangle_{1\rm{H}}|\Phi_2(\alpha_2)\rangle_{2\rm{H}}
|\Phi_3(\alpha_3)\rangle_{3\rm{H}}
\nonumber\\
&=&-\int {d\alpha'_1\over 2\pi}{d\alpha'_2\over 2\pi}
{d\alpha'_3\over 2\pi}\,
{}_1\langle\alpha'_1|\,{}_2\langle\alpha'_2|\,{}_3\langle\alpha'_3|\,
2\pi\delta(\alpha'_1+\alpha'_2+\alpha'_3)\nonumber\\
&&~~~~~~~\times\,\langle v_3|
b_0^{-(1)}b_0^{-(2)}b_0^{-(3)}\wp^{(1)}\wp^{(2)}\wp^{(3)}
|\Phi_1(\alpha_1)\rangle_{1\rm{L}}|\Phi_2(\alpha_2)\rangle_{2\rm{L}}
|\Phi_3(\alpha_3)\rangle_{3\rm{L}}\nonumber\\
&=&-2\pi\delta(\alpha_1+\alpha_2+\alpha_3)(-1)^{|\Phi_2|}
\left\langle h_1[b_0^{-}\wp\Phi_1]\,
h_2[b_0^{-}\wp\Phi_2]\,
h_3[b_0^{-}\wp\Phi_3]\,
\right\rangle.
\end{eqnarray}

%%%%%%%%%%%%%%%%%%%%%%%%%%%%%%%%%%%%%%%%%%%%%%%%%%%%%%%%%%%%%%
\section{Computation of the Determinant}
\label{sec_determinant}
%%%%%%%%%%%%%%%%%%%%%%%%%%%%%%%%%%%%%%%%%%%%%%%%%%%%%%%%%%%%%%

A nontrivial divergent factor ${\cal C}$ appears
in (\ref{eq_idempotency2}).
In this 
section,
we compute it analytically using the so-called Cremmer--Gervais identity
\cite{r:C-G}.
This identity provides us a method to compute 4-string amplitudes
constructed from 3-string vertices.
Interestingly enough, in the procedure to evaluate 4-string amplitudes,
we encounter a determinant of Neumann coefficient matrices that is
identical to the factor that we find in the
computation of the left-hand side of Eq.~(\ref{eq_idempotency2}).
The Cremmer--Gervais identity allows us to regularize the divergent factor 
${\cal C}$. This regularization corresponds to introducing 
the propagation of an intermediate string in the
4-string vertex and reveals its dependence on 
$\alpha$.

The factor that we would like to consider is
\begin{eqnarray}
\label{eq:factorC}
{\cal C} &=& \mu (1,2,3)^2 (\det (1-r^2))^{- \frac{d-2}{2}}\,,
\end{eqnarray}
where $\mu(1,2,3)$ is given in Eq.~(\ref{eq:mu123}),
and the matrix $r$ is
\begin{eqnarray}
\label{eq:rmn}
r_{mn}= \frac{\beta (\beta+1) (mn)^{3/2}}{m+n} \bar f^{(3)}_m \bar f^{(3)}_n,
\quad \bar f^{(3)}_m = \frac{\Gamma (-m \beta) e^{m ( \beta \log |\beta|
- (\beta +1) \log |\beta+1|)}}{m! \Gamma (-m \beta + 1 - m)}\,,~~~~~
\end{eqnarray}
with $\beta=\alpha_1/\alpha_3,\alpha_3=-\alpha_1-\alpha_2$.\cite{r:kmw1}
We note that this matrix $r$ is given by the Neumann coefficients for
the 3-string vertex: 
$r_{mn}=\sqrt{mn}\bar{N}^{33}_{mn}(\alpha_1,\alpha_2,\alpha_3)
=:{\tilde{N}}^{33}_{mn}$.

The Cremmer--Gervais identity applied to our case\footnote{
Here, we have used equations given in Appendix C of Ref.~\citen{r:HIKKO2}
with $(\alpha_3,\alpha_4)\rightarrow (-\alpha_2,-\alpha_1)$
and $\theta=0$. Our ${\cal A}$ corresponds to the quantity $a+c$ there.
} (which is a scattering process, such as that depicted in 
Fig.~\ref{fig:CG}) is
\begin{eqnarray}
&& \frac{1}{|\alpha_3|} \left( \det \left(1 - \tilde N^{33} \tilde N^{33}_T
\right ) \right)^{-12} e^{\frac{T}{-\alpha_3}} %\\
=\exp \left( - \log \left( Z_3^2 \frac{\partial T}{\partial Z_3} \right)
- \alpha_3^2 {\cal A} + \frac{2\tau_0}{\alpha_3} \right).~~~~~
\label{eq:cremmer-gervais}
\end{eqnarray}
\begin{figure}[htbp]
         \begin{center}
         \scalebox{0.5}[0.5]{\includegraphics{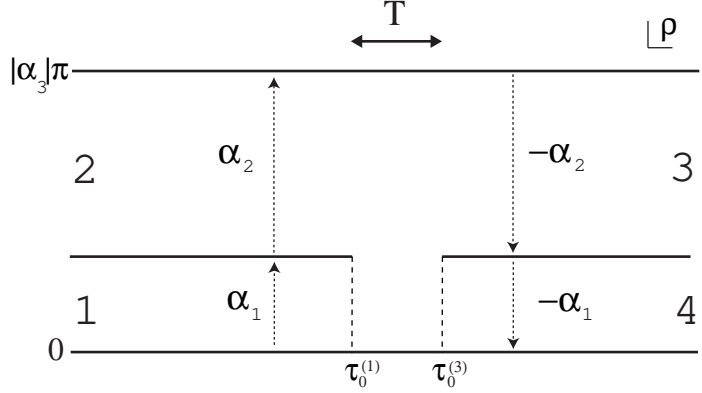}}
         \end{center}
         \caption{The 4-string configuration to obtain our determinant.
Here, we display only the ${\rm Im}\,\rho\ge 0$ region for simplicity. 
The interval $T$ corresponds to our regularization parameter.
We consider the case in which $\alpha_1,\alpha_2>0$.}
         \label{fig:CG}
\end{figure}
On the left-hand side, the matrix $\tilde{N}^{33}_T$ is given by the
Neumann coefficient for the 3-string vertex:
\begin{eqnarray}
 \tilde{N}^{33}_{Tmn}
&=&\sqrt{mn}\bar{N}^{33}_{mn}(\alpha_1,\alpha_2,\alpha_3)\,
e^{(m+n)T\over \alpha_3}\nonumber\\
&=&(-1)^{m+n}\sqrt{mn}\bar{N}^{33}_{mn}(-\alpha_2,-\alpha_1,-\alpha_3)\,
e^{(m+n)T\over \alpha_3}.
\end{eqnarray}
Here, $\alpha_3=-\alpha_1-\alpha_2$,
and $\tilde{N}^{33}_{mn}$ is the $T\rightarrow 0$ limit of
$\tilde{N}^{33}_{Tmn}$.
On the right-hand side of Eq.~(\ref{eq:cremmer-gervais}), we have
$\tau_0=\alpha_1 \log | \alpha_1 | + \alpha_2 \log |\alpha_2|
+\alpha_3\log|\alpha_3|$, and the quantity ${\cal A}$ is given by the
Neumann coefficients for the 4-string 
vertex that corresponds to Fig.~\ref{fig:CG}:
\begin{eqnarray}
\alpha_3^2 {\cal A} &=& \mbox{Re}
\biggl( \sum_{r=1}^4 \bar N_{00}^{(4)rr} - ( \bar N_{00}^{(4)12}
+ \bar N_{00}^{(4)21} + \bar N_{00}^{(4)34} + \bar N_{00}^{(4)43})
\nonumber\\
&&~~~~~~~
- 2 \tau_0 \left( \frac{1}{\alpha_1} + \frac{1}{\alpha_2} \right)
\biggr), \\
&& {\rm Re}\, \bar N_{00}^{(4)rs} = \left\{ 
\begin{array}[tb]{cc}
\log |Z_r-Z_s|\,, &  r \neq s, \\
- \sum_{i \neq r} \frac{\alpha_i}{\alpha_r} \log |Z_r - Z_i|
+ \frac{1}{\alpha_r}\tau_0^{(r)},
& r=s,
\end{array}
\right.\\
&&\tau_0^{(1)}=\tau_0^{(2)}={\rm Re}\,\rho(z_-)\,,~~
\tau_0^{(3)}=\tau_0^{(4)}={\rm Re}\,\rho(z_+)\,.
\end{eqnarray}
Here, the interaction points $z_{\pm}$ are the two solutions of 
$\frac{d \rho (z)}{d z} =0$,
where $\rho (z)$ is the Mandelstam mapping defined as
\begin{equation}
\rho (z) =\alpha_1 \log (z - Z_1)+\alpha_2 \log (z - Z_2)
-\alpha_2 \log (z - Z_3)-\alpha_1 \log (z - Z_4)\,.
\end{equation}
We fix the gauge as $Z_1=\infty,Z_2=1>Z_3>Z_4=0$. Then we obtain
\begin{eqnarray}
z_\pm &=& - (2 \alpha_1)^{-1} \left( \alpha_3 + (\alpha_2 - \alpha_1) Z_3
\pm \Delta^{\frac{1}{2}}\right)\,, 
\label{eq:zpm}\\
\Delta &=& 
( \alpha_1 + \alpha_2)^2 (1 - Z_3 ) \left\{ 1 - (2 \beta + 1)^2
Z_3 \right\}\,,
\label{eq:delta}
\end{eqnarray}
with $-1<\beta=\alpha_1/\alpha_3<0$.
The time interval $T$ that represents the propagation of 
the intermediate string is given by
\begin{eqnarray}
\label{eq:Tdef}
 T=\tau_0^{(3)}-\tau_0^{(1)}=\mbox{Re}\, ( \rho (z_+) - \rho (z_-) )\,,
\end{eqnarray}
and it is a function of $Z_3$ through $z_{\pm}$.

Now, we regularize the factor ${\cal C}$, or $r$ with the parameter $T$
so that we use Eq.~(\ref{eq:cremmer-gervais}),
\begin{equation}
(r^2)_{nm} \rightarrow (rr_T)_{nm}, \qquad (r_T)_{nm}: = e^{-(n+m)
\frac{T}{\alpha_1 + \alpha_2}} r_{nm}\,.
\end{equation}
Plugging this into Eq.~(\ref{eq:cremmer-gervais}),
we obtain the regularized expression of the factor:
\begin{eqnarray}
&&  \mu (1,2,3)^2( \det (1 - r r_T))^{-12} \nonumber \\
& =& \exp \biggl( \frac{\beta^2 + \beta + 1}{\beta ( \beta + 1)}
\frac{T}{\alpha_3} - ( 1 + \beta + \beta^2) \log Z_3\nonumber\\
&&~~~~~~~~
-\frac{1}{2} \log ( 1 - (2 \beta + 1)^2 Z_3) - \frac{3}{2} \log ( 1 - Z_3)
\biggr).~~~~
\label{eq:cg our case}
\end{eqnarray}
The right-hand side of this expression gives the factor (\ref{eq:factorC})
 in the limit $T\rightarrow 0$ for the critical dimension, $d=26$.
It can be seen from Eqs.~(\ref{eq:Tdef}), (\ref{eq:zpm}) and
 (\ref{eq:delta}) that this corresponds to taking $z_+ \rightarrow z_-$,
and consequently $Z_3 \rightarrow 1$.
If we define $\varepsilon \equiv 1 - Z_3$, then $T$ can be expressed as
\begin{equation}
\frac{T}{|\alpha_3|} = 4 \varepsilon^{\frac{1}{2}} \sqrt{- \beta ( \beta +1)}
+ {\cal O} (\varepsilon)\,.
\end{equation}
In the limit that $Z_3\rightarrow 1$, Eq.~(\ref{eq:cg our case}) reduces to
\begin{eqnarray}
 \mu (1,2,3)^2( \det ( 1 - r r_T))^{-12}
&=& 2^5
\left( \frac{T}{|\alpha_3|} \right)^{-3} \left[ - \beta ( \beta+1) \right]
\left\{ 1 + {\cal O} (T) \right\} \nonumber\\
&\rightarrow & 2^5 T^{-3} |\alpha_1 \alpha_2 (\alpha_1+\alpha_2)|\,.
\label{eq:alphadep}
\end{eqnarray}

The regularization that we adopted here for evaluation of the factor 
${\cal C}$, i.e. Eq.~(\ref{eq:factorC}), is consistent with 
the level truncation approximation if we make the identification
$T^{-1}\sim L$, where $L$ is the
size of the truncated matrix $r$ (\ref{eq:rmn}), 
or Neumann matrix $\bar{N}^{33}$.
Although we observed that ${\cal C}\propto L^3$ in Ref.~\citen{r:kmw1},
we further investigated its $\beta$ dependence in the case of the
critical dimension, $d=26$.
In fact, we have plotted ${\cal C}/(L^3(-\beta)(\beta+1))$ 
up to $L=2000$ using the computer program
{\sl Mathematica5} and 
confirmed its convergence to a constant ($\sim 772$)
independent of $\beta$ (see Fig.~\ref{fig:4figs_v1}).
This numerical result implies that 
our regularization of ${\cal C}$, Eq.~(\ref{eq:alphadep}),
is consistent with the level truncation 
of $\bar{N}^{33}_{mn}(\alpha_1,\alpha_2,\alpha_3)$
through the identification of the parameters $L\sim {|\alpha_3|/T}$\,.
\begin{figure}[htbp]
         \begin{center}
         \scalebox{0.7}[0.7]{\includegraphics{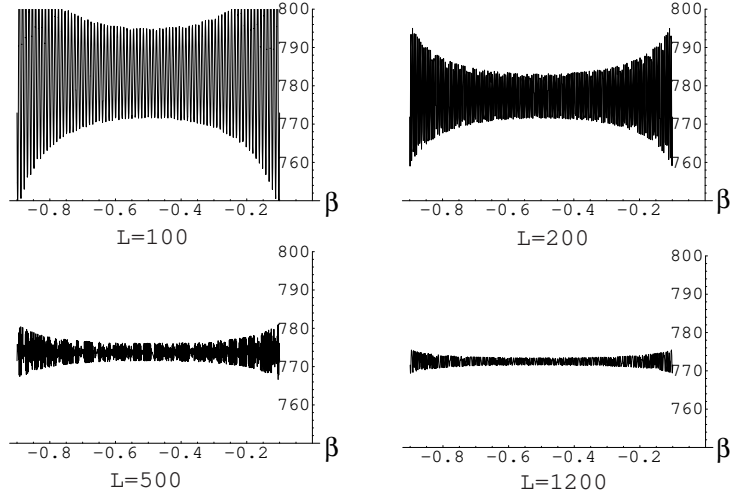}}
         \end{center}
         \caption{Plots of ${\cal C}/(L^3(-\beta)(\beta+1))$ 
for $L=100,200,500,1200$ 
in the interval $-0.9\le \beta\le -0.1$ obtained using
{\sl Mathematica5}.
Although this quantity oscillates as a function of $\beta$
for small $L$, it converges to a constant
independent of $\beta$ as $L\rightarrow \infty$.}
         \label{fig:4figs_v1}
\end{figure}

\end{document}